\documentclass
[aps,prb,twocolumn,floatfix,english,showpacs,10pt]{revtex4-2}%
\usepackage{graphicx}
\usepackage{booktabs}
\usepackage{amsmath}
\usepackage{physics}
\usepackage{amssymb}
\usepackage{colordvi}
\usepackage{verbatim}
\usepackage{xcolor}
\usepackage{mathrsfs}
\usepackage{epsfig}
\usepackage{lipsum}
\usepackage{amsfonts}
\usepackage{makecell}
\usepackage{esint}
\usepackage{bm}
\usepackage[most]{tcolorbox}

\usepackage[unicode=true, breaklinks=false, pdfborder={0 0 1}, backref=false,
colorlinks=true, linkcolor=blue, urlcolor=blue, citecolor=blue]{hyperref}%
\providecommand{\U}[1]{\protect\rule{.1in}{.1in}}
\providecommand{\U}[1]{\protect\rule{.1in}{.1in}}
\setcitestyle{numbers,square}

\begin{document}

\title{Pumping of spin supercurrent in unitary triplet superconductors}

\author{Ping Li}
\affiliation{School of Physics, Huazhong University of Science and Technology, Wuhan 430074, China}

\author{Tao Yu}
\email{taoyuphy@hust.edu.cn}
\affiliation{School of Physics, Huazhong University of Science and Technology, Wuhan 430074, China}

\date{\today }

\begin{abstract}

One efficient mechanism for generating a charge supercurrent is Andreev reflection, in which the electric current injected from a normal metal into a conventional superconductor is converted into a supercurrent, thereby preserving charge conservation. We here propose a general principle for generating spin supercurrents in triplet superconductors by analogy with such charge transport, i.e., assuming spin conservation. We find a spin torque that is proportional to the triplet superconducting order parameter and, in the spin-conservation scenario, converts the particle spin to that of Cooper pairs. Based on this general principle, we propose an implementation to efficiently generate a spin supercurrent in unitary triplet superconductors, even though Cooper pairs carry no spin polarization at equilibrium, by the magnetization dynamics ${\bf M}(t)$ of a proximity magnetic nanostructure. The efficiency of this spin pumping is not solely limited to the $d{\bf M}/dt\times {\bf M}$ due to the emergent particle-hole symmetry, thereby going beyond the conventional spin pumping of electrons. This general principle provides an efficient approach to generating and manipulating dissipationless spin currents in many unconventional superconductors.

\end{abstract}

\maketitle

\section{Introduction}

Spin supercurrent refers to the flow of spin angular momentum that is dissipationless in transport, which is an excellent property that has been pursued in spintronic devices for decades.
Its realization may rely on the triplet superconductivity, described by the order parameter $\Delta({\bf k})=[{\bf d}({\bf k})\cdot{\pmb \sigma}]i\sigma_y$ in different disguises, where ${\bf k}$ is the wave vector and ${\pmb \sigma}$ is the Pauli matrices. 
The ferromagnet$|$superconductor heterostructures~\cite{JLIN,review_SOC,FSBer,AIBuz,JLind,KRjeon,KRjeon1} exploit the interface between conventional superconductors and  ferromagnets~\cite{RGrein,Ingvild,GABo,JLinder1,KHalterman}, magnetic textures~\cite{TChampel,MAlidoust,Zahra,ADBer,SammeM}, or altermagnets~\cite{Hans} to break the spin degeneracy and specific spin-rotation symmetries, which may induce the triplet Cooper pairing. 
Spin-orbit coupling may favor the formation of long-range triplet proximity effect in these heterostructures~\cite{FS-SOC,FS-LR,review_SOC,JLinder,XMontiel,SHJacobsen,SOC_TY} and in conventional superconductors, rendering a triplet Cooper pairing with its ${\bf d}({\bf k})$-vector aligned to the spin polarization of electrons~\cite{MSigrist,PAFrigeri,EB-HeavyFermion,MS-non-centro,EB-Non-centro,Nagaosa_SOC}. Intrinsic triplet superconductors are useful for spin transport and quantum computing; experiments have already accumulated evidences for several candidates, \textit{e.g.}, $\rm{K_2Cr_3As_3}$~\cite{JYang,JKBao,KMTaddei,Huakunzuo}, $\rm{UTe_2}$~\cite{ShengRan,TMetz,DAoki}, $\rm{Sr_2RuO_4}$~\cite{AP} (recent investigations suggest $\rm{Sr_2RuO_4}$ may not be a $p$-wave superconductor~\cite{not_1,not_2,not_3}), $\rm{UPt_3}$~\cite{JD,HTou}, $\rm Cu_xBi_2Se_3$~\cite{MKriener,MYokoyama}, and LaAlO$_3$/KTaO$_3$ superconducting interface~\cite{interface,C_Liu,Z_Chen,Hanwei}. Studies have showed that when the Cooper pairs carry spin polarization with condensate spin polarization ${\rm Tr}(\Delta^{\dagger}({\bf k}){\pmb \sigma}\Delta({\bf k}))=-i{\bf d}({\bf k})\times {\bf d}^*({\bf k})\ne 0$, i.e., the triplet superconductivity is non-unitary,  the system may transport the spin supercurrent by a formation of center-of-mass momentum under proper bias, \textit{e.g.}, by the temperature gradient~\cite{supercurrent_Seebeck} or the charge supercurrent~\cite{GABo}. A natural question arises whether a triplet superconductor without carrying the intrinsic Cooper-pair polarization ${\rm Tr}(\Delta^{\dagger}({\bf k}){\pmb \sigma}\Delta({\bf k}))=-i{\bf d}({\bf k})\times {\bf d}^*({\bf k})= 0$ supports a spin supercurrent and how to drive it.

\begin{figure}[htp!]	
\includegraphics[width=0.57\textwidth,trim=1.5cm 0cm 0cm 0.0cm]{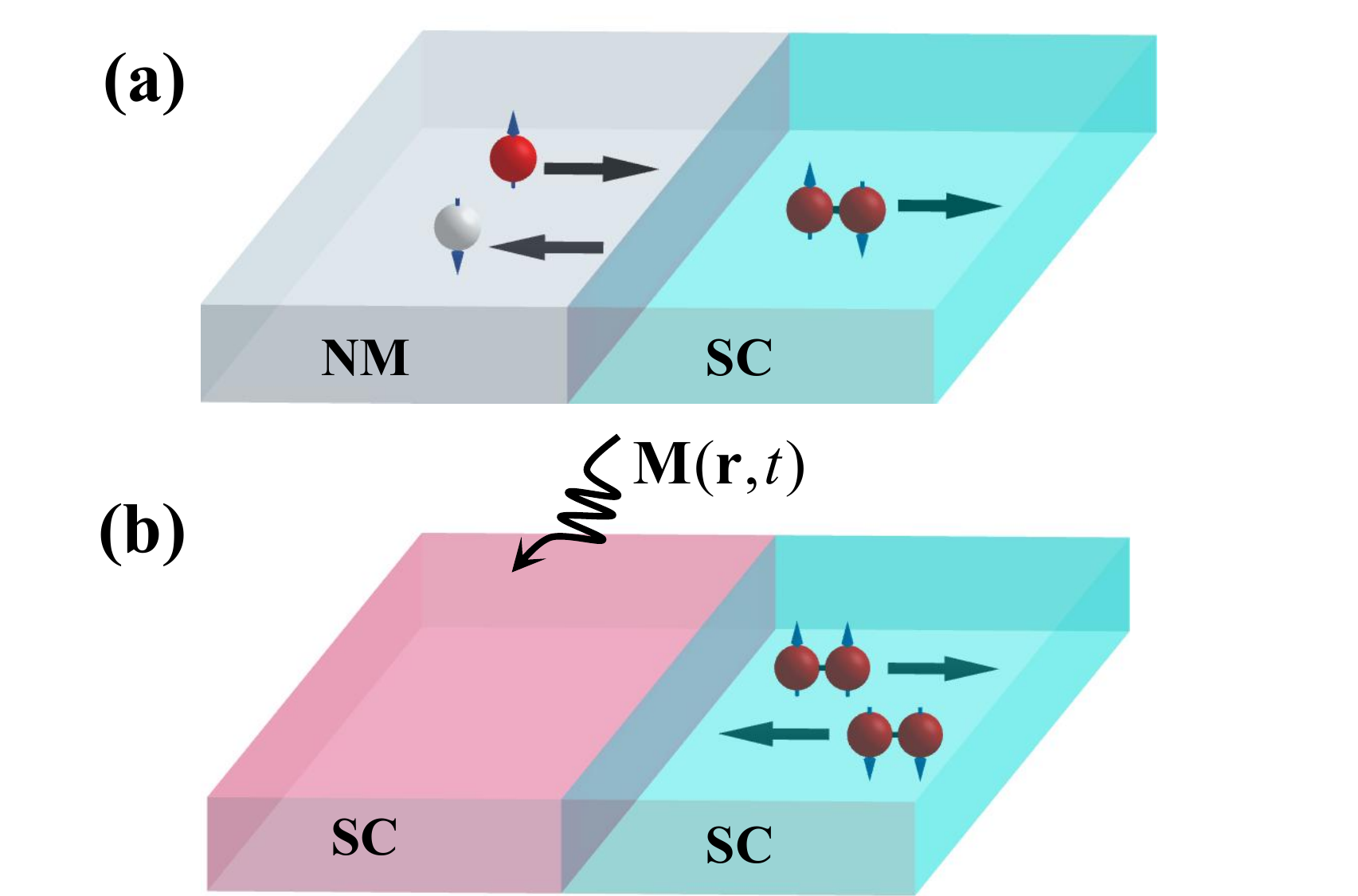}
\caption{Comparison of the generation of charge supercurrent in conventional $s$-wave superconductors [(a)] and the pumping of spin supercurrent in triplet $p$-wave superconductors with unitary triplet Cooper pairing [(b)]. In (a), the electric current injected from a normal metal (NM) to the conventional superconductor (SC) converts to a supercurrent carried by Cooper pairs via the Andreev reflection by retaining charge conservation. In (b), the injection of spin torque from the driven region by a local dynamical exchange field to the non-driven region generates a spin supercurrent by retaining the spin conservation.}
\label{model}
\end{figure}

In this work, we answer this question affirmatively based on a general principle of spin conservation in a prototypical $p$-wave superconductor with a unitary triplet superconductivity, via making an analogy to charge conservation in the Andreev reflection~\cite{Andreev_reflection}. 
The charge carried by \textit{quasiparticles} itself in superconductors is not conserved due to the $U(1)$ symmetry breaking, which is restored by including the charges carried by the Cooper-pair condensate~\cite{charge_conservation_1,charge_conservation_2,charge_conservation_3,charge_conservation_4,Tinkham_book,BTK}. Indeed, using the wavefunction of quasiparticle \(\Psi = (f, g)\),  the charge density of quasiparticles \( Q = e(|f|^2 - |g|^2) \) is not conserved in the continuity equation $ \partial_t Q + \nabla \cdot \mathbf{J}_Q = 4e\Delta_s  \text{Im}(f^* g)$ due to the $s$-wave superconducting order parameter $\Delta_s$, in which the charge current carried by quasiparticles $ \mathbf{J}_Q = ({e}/{m}) \left[ \text{Im}(f^* \nabla f) + \text{Im}(g^* \nabla g) \right]$. The charge sink \( 4e\Delta_s \, \text{Im}(f^* g) \) is then interpreted as a source of the superconducting condensate~\cite{Tinkham_book,BTK,Andreev_reflection}, which renders the generation of charge supercurrent in Andreev reflection in which the refection of the particle to antiparticle converts to a Cooper pair~\cite{Andreev_reflection} as in Fig.~\ref{model}(a). An analogous understanding of the generation of spin supercurrents has not, to the best of our knowledge, been addressed in unconventional triplet superconductors.

We generalize this scenario of charge supercurrent generation in conventional superconductors to the pumping of a spin supercurrent in triplet superconductors with unitary Cooper pairing.
We demonstrate that, similar to the singlet $s$-wave superconducting order parameter that acts as the charge sink~\cite{Tinkham_book,BTK,Andreev_reflection}, the triplet superconducting order parameter renders a spin sink or a spin torque to the injected spin polarization carried by the quasiparticles, which, by restoring the spin conservation, is converted to a dissipationless spin supercurrent. 
Based on this general principle, we propose generating spin supercurrent through spin pumping driven by \textit{local} magnetization ${\bf M}(t)$ dynamics of proximity magnetic nanostructures. 
As illustrated in Fig.~\ref{model}(b), a local exchange magnetic field separates the superconductor film into the ``dynamic" and ``static" parts (with a ``smooth" interface), which generates a spin torque in the driven region that pumps a spin supercurrent from the former to the latter. In this spin-pumping mechanism, the generated spin supercurrent is no longer limited to the conventional form $\propto (d{\bf M}/dt)\times {\bf M}$, thereby going beyond the conventional spin pumping~\cite{spin_pumping_1}. Implementing the general principle with these efficient mechanisms provides a pathway to generate and manipulate dissipationless spin currents in unconventional superconductors.

We organize this article as follows. We first demonstrate universally in the $p$-wave superconductor that the triplet order parameter renders a spin torque to the spin polarization of quasiparticles and derive the associated spin-current operator by the continuity equation for the spin density in Sec.~\ref {spin_torque}. In Sec.~\ref{exchange_interaction} and \ref{scattering_theory}, we develop a general scattering theory to calculate the spin-current and spin-torque densities when the local time-dependent external fields drive the triplet superconductors. Section~\ref{spin_pumping}  provides the typical realizations of the pumping of dissipative and dissipationless spin current by the local magnetization dynamics in the adjacent magnetic nanostructure. We discuss and conclude our results in Sec.~\ref{discussion_conclusion}. Appendix~\ref{spin_current_appendix} and \ref{spin_torque_appendix} provide details in the derivation of spin-current and spin-torque densities.

\section{Spin torque by triplet order parameters}
\label{spin_torque}

We start by addressing the conservation of spin density $\hat{\bf s}({\bf r})$ in the triplet $p$-wave superconductors by defining the associated spin-current density $\hat{\bf J}_s({\bf r})$ and spin-torque density $\hat{\bf T}_s({\bf r})$ operators.

We illustrate the general principle by considering a thin superconductor film with low electron densities or two-dimensional electron gas with the surface normal along the $\hat{\bf x}$-direction, with which ${\bf r}=(x=0,\pmb{\rho})$ with the in-plane ${\pmb \rho}=y\hat{\bf y}+z\hat{\bf z}$. 
In terms of the field operator $\hat{\Psi}({\pmb \rho})=\left(\hat{\psi}_\uparrow(\pmb \rho),\hat{ \psi}_\downarrow(\pmb \rho),\hat{\psi}^\dagger_\uparrow(\pmb \rho),\hat{ \psi}^\dagger_\downarrow(\pmb \rho)\right)^T$ in the Nambu space, the Hamiltonian $\hat{H}_0=(1/2)\int d{\pmb \rho}\hat{\Psi}^\dagger({\pmb \rho}){\cal H}_0({\pmb \rho})\hat{\Psi}({\pmb \rho})$ for a prototypical triplet $p$-wave superconductor is described by the Hamiltonian matrix 
\begin{align}
{\cal H}_0({\pmb \rho})=
\left(\begin{array}{cccc}
\frac{\hat{\bf p}^2}{2m}-\mu & 0 & \Delta_p e^{i \theta_{\bf k}} & 0 \\
0 & \frac{\hat{\bf p}^2}{2m}-\mu & 0 & -\Delta_p e^{-i\theta_{\bf k}} \\
\Delta_p e^{-i\theta_{\bf k}}& 0 & -(\frac{\hat{\bf p}^2}{2m}-\mu) & 0 \\
0 & -\Delta_p e^{i\theta_{\bf k}} & 0 & -(\frac{\hat{\bf p}^2}{2m}-\mu)\end{array}
\right).
\label{Hamiltonian}
\end{align}
It contains the diagonal kinetic energy with respect to the electron chemical potential $\mu$, in which $m$ is the electron mass and $\hat{p}_{y,z}=\hbar \hat{k}_{y,z}=-i\hbar \nabla_{y,z}$ is the momentum operator, and the off-diagonal triplet superconductivity governed by the order parameter $\Delta_p$ and $e^{i\theta_{\bf k}}=(\hat{k}_y+i\hat{k}_z)/|{\bf k}|$. This type of triplet superconductivity was recently proposed to explain the Andreev bound states and other properties observed in LaAlO$_3$/KTaO$_3$ superconducting interface~\cite{interface,C_Liu,Z_Chen,Hanwei}. In such triplet superconductors,  
\[
\left(\begin{array}{cc}
 \Delta_p e^{i \theta_{\bf k}} & 0 \\
  0 & -\Delta_p e^{-i\theta_{\bf k}} 
\end{array}
\right)=[{\bf d}({\bf k})\cdot{\pmb \sigma}]i\sigma_y
\]
is unitary since, with ${\bf d}({\bf k})=\Delta_p(-k_y/\abs{\bf k},{k}_z/{\abs{\bf k}},0)$, $i({\bf d}^*({\bf k})\times {\bf d}({\bf k}))=0$, \textit{i.e.}, the Cooper pairs carry no spin polarization at equilibrium.

The spin density operator $\hat{\bf s}({\pmb \rho})=(\hbar/4)\hat{\Psi}^{\dagger}({\pmb \rho}){\pmb{\cal S}}\hat{\Psi}({\pmb \rho})$, 
in which  
${\pmb{\cal S}}=\left(\begin{array}{cc}{ \pmb\sigma}  &  \\
& -{\pmb\sigma}^\ast
\end{array}\right)$ 
represents the Pauli matrices in the Nambu space, evolves according to the continuity equation
\begin{align}
\partial_t\hat{\bf s}^\nu({\pmb \rho})+\pmb\nabla\cdot\hat{\bf J}_s^\nu({\pmb \rho})=\hat{\bf T}^{\nu}_s({\pmb \rho}),
\label{spin_continuity_equation}
\end{align}
which defines the spin-current density operator
\begin{align}
    \hat{\bf J}_{ s}({\pmb \rho})=\frac{i\hbar^2}{8m}\hat{\Psi}^\dagger({\pmb \rho})(\overleftarrow{\pmb{\nabla}}-\overrightarrow{\pmb{\nabla}})\tau_3{\pmb{\cal S}}\hat{\Psi}({\pmb \rho}),
    \label{spin_current_operator}
\end{align}
where $\tau_3={\rm diag}(1,1,-1,-1)$ is the metric in the Nambu space and the arrows ``$\rightarrow$" and ``$\leftarrow$" indicate the direction to which the operator acts. 
The spin density $\hat{\bf s}({\pmb \rho})$ of quasiparticles is then not conserved in the presence of the triplet order parameter $\Delta_p$ that renders a spin-torque density operator
\begin{align}
    \hat{\mathbf{T}}_s({\pmb \rho})&=\frac{\Delta_p}{4i}\hat{\Psi}^\dagger({\pmb \rho})\Big[\left(\begin{array}{cccc}
         0&0&{\overleftarrow{e^{i\theta_{\bf k}}}}&0  \\
         0&0&0&-\overleftarrow{e^{-i\theta_{\bf k}}}\\
         \overleftarrow{e^{-i\theta_{\bf k}}}&0&0&0\\
         0&-\overleftarrow{e^{i\theta_{\bf k}}}&0&0
    \end{array}\right){\pmb{\cal S}}\nonumber\\
    &+{\pmb{\cal S}}\left(\begin{array}{cccc}
         0&0&{\overrightarrow{e^{i\theta_{\bf k}}}}&0\\
         0&0&0&-\overrightarrow{e^{-i\theta_{\bf k}}}\\
         \overrightarrow{e^{-i\theta_{\bf k}}}&0&0&0\\
         0&-\overrightarrow{e^{i\theta_{\bf k}}}&0&0
    \end{array}\right)\Big]\hat{\Psi}({\pmb \rho}).
    \label{spin_torque_density_operator}
\end{align}

\begin{figure}[htp!]	
\includegraphics[width=0.48\textwidth,trim=0cm 0cm 0cm 0.0cm]{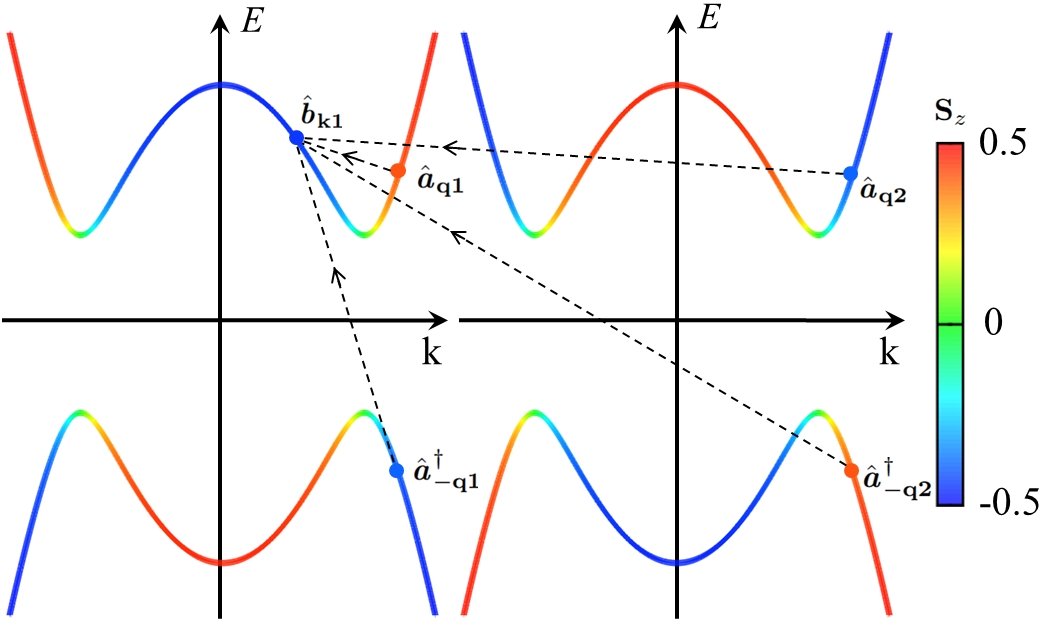}	
\caption{Spin of quasiparticles and illustration of scattering process with quasiparticles scattered from $\{\hat{a}_{{\bf q}1},\hat{a}_{{\bf q}2},\hat{a}_{{\bf q}3},\hat{a}_{{\bf q}4}\}$ to $\hat{b}_{{\bf k}1}$ that includes the intraband and interband processes. The color map on the band indicates the spin of quasiparticles $S_z\in[-\hbar/2,\hbar/2]$.}
\label{scattering_process}
\end{figure}

According to the Hamiltonian matrix~\eqref{Hamiltonian}, the dispersion of quasiparticles $E_{1}(k)=E_{2}(k)=\sqrt{\xi_k^2+\Delta_p^2}\equiv E_k$ and $E_{3}(k)=E_{4}(k)=-\sqrt{\xi_k^2+\Delta_p^2}\equiv -E_k$ with $\xi_k={\hbar^2k^2}/({2m})-\mu$ are degenerate in spin. The frequencies of quasiparticles are then $\omega_{\alpha=1,2,3,4}(k)=E_{\alpha}(k)/\hbar$. Their eigenstates 
\begin{align}
     &\phi_{\alpha=1}({\bf k})=\left(\begin{array}{c}
		     u_{1,\uparrow}({\bf k})  \\
		     u_{1,\downarrow}({\bf k})\\
       v_{1,\uparrow}({\bf k})\\
       v_{1,\downarrow}({\bf k})
		\end{array}\right)=\left(\begin{array}{c}
\sqrt{\frac{E_k+\xi_k}{2E_k}} \\
0\\
\frac{\Delta_pe^{-i\theta_{\bf k}}}{\sqrt{2E_k(E_k+\xi_k)}}\\
    0
\end{array}\right),\nonumber
\end{align}
\begin{align}
&\phi_{\alpha=2}({\bf k})=\left(\begin{array}{c}
u_{2,\uparrow}({\bf k})  \\
u_{2,\downarrow}({\bf k})\\
v_{2,\uparrow}({\bf k})\\
v_{2,\downarrow}({\bf k})
\end{array}\right)=\left(\begin{array}{c}
0\\
\sqrt{\frac{E_k+\xi_k}{2E_k}} \\
0\\
\frac{-\Delta_pe^{i\theta_{\bf k}}}
{\sqrt{2E_k(E_k+\xi_k)}}
\end{array}\right),\nonumber
\end{align}
\begin{align}
    &\phi_{\alpha=3}({\bf k})=\left(\begin{array}{c}
		     u_{3,\uparrow}({\bf k})  \\
		     u_{3,\downarrow}({\bf k})\\
       v_{3,\uparrow}({\bf k})\\
       v_{3,\downarrow}({\bf k})
		\end{array}\right)=\left(\begin{array}{c}
     -\frac{\Delta_pe^{i\theta_{\bf k}}}{\sqrt{2E_k(E_k+\xi_k)}}\\
    0\\ \sqrt{\frac{E_k+\xi_k}{2E_k}} \\
		    0
		\end{array}\right),\nonumber
    \end{align}
    \begin{align}
        &\phi_{\alpha=4}({\bf k})=\left(\begin{array}{c}
		     u_{4,\uparrow}({\bf k})  \\
		     u_{4,\downarrow}({\bf k})\\
       v_{4,\uparrow}({\bf k})\\
       v_{4,\downarrow}({\bf k})
		\end{array}\right)=\left(\begin{array}{c}
      0\\
     \frac{\Delta_pe^{-i\theta_{\bf k}}}{\sqrt{2E_k(E_k+\xi_k)}}\\
    0\\
\sqrt{\frac{E_k+\xi_k}{2E_k}} 
\end{array}\right) .
\label{eigenstates}
 \end{align}
In the presence of the triplet Cooper pairing, the amplitudes of spin ``$\uparrow$" and spin ``$\downarrow$" in the eigenstates, although not mixed, become wave-vector dependent. This renders the spin carried by quasiparticles in different bands wave-vector-dependent, a feature reflected in the color map in Fig.~\ref{scattering_process}. It is then substantiated that the spin is not conserved for the quasiparticles themselves during the scattering process.

\section{Exchange interaction with magnetization}
\label{exchange_interaction}

Generally, we consider here the coupling Hamiltonian between the ferromagnetic nanostructures and the electron spins due to the interfacial exchange interaction
\begin{align}
\hat{V}&=J_{\rm ex}\int\hat{\mathbf{s}}(\pmb{\rho},t)\cdot
{\mathbf{M}}(x=0,\pmb{\rho},t)d\pmb{\rho}\nonumber\\
&=\frac{J_{\rm ex}\hbar}{4}\int d{\boldsymbol\rho}\hat{\Psi}^\dagger({\boldsymbol\rho}){\pmb{\cal S}}\cdot{\mathbf  M}(x=0,{\pmb \rho},t)\hat{\Psi}(\pmb{\rho}),
\label{exchange_Hamiltonian}
\end{align}
where $J_{\rm ex}$ is the exchange coupling constant that induces the local exchange splitting $\Delta= J_{\rm ex}\hbar M_s$ of electrons by the ferromagnet with a saturation magnetization $M_s$. 
Here, the energy splitting is local to the $O(100)$~nm region. This region is much shorter than the coherent length of the superconductor; the associated Zeeman splitting is much smaller than the condensation energy of electrons. Thereby, we assume the local exchange field does not affect the overall superconductivity. 
We therefore focus on the dynamical response of the superconductor to the coherent magnetization dynamics.

In the ferromagnetic nanostructures, the coherent magnetization $\mathbf{M}({\bf r},t)$ precesses around the saturation magnetization. The AC exchange magnetic field due to the exchange coupling \eqref{exchange_Hamiltonian} 
\begin{align}
    {\bf H}_{\rm ex}(x=0,{\pmb \rho},t)&\equiv ({J_{\rm ex}\hbar}/{4}){\bf M}(x=0,{\pmb \rho},t)\nonumber\\
    &=\frac{1}{A}\sum_{\xi=\pm }\sum_{\bf k}{\bf H}_{\rm ex}^{\xi}({\bf k})e^{i{\bf k}\cdot {\pmb \rho}-i\xi \omega t}
\end{align}
oscillates by the frequency $\omega$. For a two-dimensional superconductor with an area $A$, the matrix ${U}({\bf k}) =(\phi_1({\bf k}),\phi_2({\bf k}),\phi_3({\bf k}),\phi_4({\bf k}))$ and annihilation operators of quasiparticles $\hat{\textbf{\emph a}}_{\bf k}=(\hat{a}_{{\bf k}1},\hat{a}_{{\bf k}2} ,
\hat{a}^\dagger_{-{\bf k}1},\hat{a}^\dagger_{-{\bf k}2})^T$ expand the field operator in terms of quasiparticles
\begin{align}
    \hat{\Psi}({\pmb \rho})=\frac{1}{\sqrt{A}}\sum_{\bf k}{U}({\bf k}) e^{i{\bf k}\cdot{\pmb \rho}}\hat{\textbf{\emph a}}_{\bf k}.
    \label{field_operator}
\end{align}
 We note that by particle-hole symmetry, $\hat{a}_{{\bf k}3}=\hat{a}^\dagger_{-{\bf k}1}$ and $\hat{a}_{{\bf k}4}=\hat{a}^\dagger_{-{\bf k}2}$.
In terms of the field operator \eqref{field_operator}, the interaction Hamiltonian \eqref{exchange_Hamiltonian} due to the interfacial exchange coupling is then written as 
\begin{align}
     \hat{V}(t)=\sum_{{\bf k} {\bf k}^\prime}\sum_{\xi=\pm}\sum_{\alpha,\beta=1}^4{\mathcal{G}^\xi_{\alpha\beta}({\bf k},{\bf k}^\prime)}\hat{a}^\dagger_{{\bf k}\alpha}\hat{a}_{{\bf k}^\prime\beta}e^{-i\xi\omega t},
     \label{interection_Hamiltonian}
 \end{align}
 where the $4\times4$ coupling matrix in Eq.~\eqref{interection_Hamiltonian} 
\begin{align}
  \mathcal{G}^\xi({\bf k},{\bf k}^\prime)&=(1/A){\bf H}^\xi_{\rm ex}({\bf k}-{\bf k}^\prime)\cdot U^\dagger({\bf k})\pmb{\cal S}U({\bf k}^\prime)
  \label{coupling_constants_ex}
\end{align}
are the coupling constants between quasiparticles of wave vector ${\bf k}$ and ${\bf k}'$ mediated by the external fields.

\section{Scattering theory}
\label{scattering_theory}

We formulate the nonlinear dynamics of triplet superconductors in terms of the scattering theory. 
When a local AC exchange field is applied to the superconductor, the external driven field \eqref{interection_Hamiltonian} acts as the local time-dependent scattering potential that scatters the incident quasiparticles. 
To this end, we develop the scattering theory to calculate the pumped spin-current density \eqref{spin_current_operator} and spin-torque density \eqref{spin_torque_density_operator}.

We derive the wave function of quasiparticles in the presence of the time-dependent scattering potential using the time-dependent perturbation theory. 
The time evolution operator $\hat{U}_I(t,t_0)=T\exp{-(i/\hbar)\int^t_{t_0}{\cal V}_I({\pmb \rho},t^\prime)dt^\prime}$ in the interaction picture ``$I$",
in which $T$ stands for the time-ordering operator and ${\cal V}_I({\pmb \rho},t)=e^{i{\cal H}_0({\pmb \rho})t/\hbar}{\cal V}({\pmb \rho})e^{-i{\cal H}_0({\pmb \rho})t/\hbar}$ is the perturbation Hamiltonian in the interaction representation.
When the interaction ${\cal V}({\pmb \rho},t)$ is introduced adiabatically from $t_0=-\infty$, the initial wave function is assumed to be a specific eigenstate $\phi_n({\pmb \rho},t_0\rightarrow-\infty)$.
According to the time evolution operator, the wave function  at time $t$ in the interaction representation reads
\begin{align}
    \Phi_I({\pmb \rho},t)=\hat{U}_I(t,t_0\rightarrow-\infty)\phi_{n,I}({\pmb \rho},t_0\rightarrow-\infty).
\end{align}
In the Schr\"odinger picture, retaining up to the second order of ${\cal V}_I({\pmb \rho},t)$, we arrive at the wave function at time $t$
 \begin{align}
     \Phi({\pmb \rho},t)&=e^{-i{\cal H}_0({\pmb \rho})t/\hbar}\hat{U}_I(t,t_0\rightarrow-\infty)\phi_n({\pmb \rho})\nonumber\\
     &=e^{-i{\cal H}_0({\pmb \rho})t/\hbar}\Big[1-\frac{i}{\hbar}\int^t_{t_0}{\cal V}_I({\pmb \rho},t^\prime)dt^\prime+\left(-\frac{i}{\hbar}\right)^2\nonumber\\
&\times\int_{t_0}^tdt^\prime\int_{t_0}^{t^{\prime}} dt^{\prime\prime}{\cal V}_I({\pmb \rho},t^\prime){\cal V}_I({\pmb \rho},t^{\prime\prime})+\dotsi\Big]\phi_n({\pmb \rho}).
\label{evolution}
\end{align}

We can, in principle, find the wave function of quasiparticles at time $t$  using the time-dependent perturbation theory to the desired orders of interaction  
\begin{align}
   \Phi_{\alpha}({\pmb \rho},t)=\sum_{{\bf q}'}\sum_{\alpha'=1}^4T_{\alpha'\alpha}({\bf q}',{\bf k},t)\phi_{\alpha'}({\bf q}')e^{i({\bf q}'\cdot{\pmb \rho}-\omega_{\alpha'}(q')t)},
\label{pumped_wave_function}
\end{align}
in which the elements of the $T$-matrix read
\begin{align}
    &T_{\alpha'\alpha}({\bf q}',{\bf k},t)=\delta_{{\bf q}'{\bf k}}\delta_{\alpha'\alpha}\nonumber\\
    &+\sum_{\xi=\pm}\Gamma_{\alpha'\alpha}^{\xi}({\bf q}',{\bf k})\exp\left[-i\left(\xi\omega-\omega_{\alpha'}(q')+\omega_\alpha(k)\right)t\right]\nonumber\\
    &+\sum_{\xi_1\xi_2}\Delta_{\alpha'\alpha}^{\xi_1\xi_2}({\bf q}',{\bf k}) \exp\left[-i(\omega_{\alpha}(k)+(\xi_1+\xi_2)\omega-\omega_{\alpha'}(q'))t\right]\nonumber\\
    &+\cdots.
\end{align}
The complex numbers 
\begin{align}
\Gamma_{\alpha'\alpha}^{\xi}({\bf q}',{\bf k})=\frac{\mathcal{G}^\xi_{\alpha'\alpha}({\bf q}',{\bf k})}{\xi\hbar\omega-E_{\alpha'}(q')+E_\alpha(k)+i\delta}
\end{align}
denote the scattering amplitudes when the electron absorbs ($\xi=-$) or emits ($\xi=+$) one magnon, where $\delta\rightarrow0^+$ is introduced due to the adiabatic introduction of the interaction, while the amplitudes  
\begin{align}
    \Delta_{\alpha'\alpha}^{\xi_1\xi_2}({\bf q}',{\bf k})&
    =\sum_{{\bf q}'',\alpha''}\frac{\mathcal{G}^{\xi_1}_{\alpha'\alpha''}({\bf q}',{\bf q}'')}{E_{\alpha''}(q'')-E_\alpha(k)-\xi_2\hbar\omega-i\delta}\nonumber\\
    &\times \frac{\mathcal{G}^{\xi_2}_{\alpha''\alpha}({\bf q}'',{\bf k})}{E_{\alpha'}(q')-E_\alpha(k)-(\xi_1+\xi_2)\hbar\omega-i\delta}
    \label{scattering_amplitudes_2}
\end{align}
are the scattering amplitudes involving the two-magnon processes. Particularly, $\{\xi_1,\xi_2\}=\{-,-\}$ represent the absorption of two magnons; $\{\xi_1,\xi_2\}=\{+,+\}$ denote the emission of two magnons; and $\{\xi_1,\xi_2\}=\{+,-\}$ or $\{-,+\}$ represent first emission and then absorption of one magnon or vice versa.

We can interpret the evolution of the wave function \eqref{pumped_wave_function} in the superconductors using the scattering theory. To this end, we define the operator
\begin{align}
    \hat{b}_{{\bf k}\alpha}=\sum_{{\bf q}'}\sum_{\alpha'=1}^4T_{\alpha\alpha'}({\bf k},{\bf q}',t)\hat{a}_{{\bf q}'\alpha'}
\end{align}
that includes the contributions of the incident operator of quasiparticles $\hat{a}_{{\bf q}'\alpha'}$
 and the scattering operator, such that $\hat{b}^{\dagger}_{{\bf k}\alpha}=\sum_{{\bf q}'\alpha'}T^*_{\alpha\alpha'}({\bf k},{\bf q}',t)\hat{a}^{\dagger}_{{\bf q}'\alpha'}=\sum_{{\bf q}'\alpha'}T_{\alpha'\alpha}({\bf q}',{\bf k},t)\hat{a}^{\dagger}_{{\bf q}'\alpha'}$.
 Here, $\hat{b}^\dagger_{{\bf k}\alpha}$ acts on the quasiparticle ground state $\ket{0}_{{Q}}$; upon taking the coordinate representation, we arrive at the wave function \eqref{pumped_wave_function} via 
\begin{align}
    \bra{\pmb \rho}\hat{b}^\dagger_{{\bf k}\alpha}\ket{0}_{{Q}}&=\sum_{{\bf q}'\alpha'} \bra{\pmb \rho}{T}^\dagger_{\alpha\alpha'}({\bf k},{\bf q}',t)\hat{a}^\dagger_{{\bf q}'\alpha'}\ket{0}_{Q}\nonumber\\
    &=\sum_{{\bf q}'\alpha'}{T}_{\alpha'\alpha}({\bf q}',{\bf k},t)\phi_{\alpha'}({\bf q}')e^{i({\bf q}'\cdot{\pmb \rho}-\omega_{\alpha'}(q')t)}.
    \nonumber
\end{align}
It indicates that the effect of the external AC  field on the electron dynamics is encoded in the operator $\hat{\pmb b}_{\bf k}(t)=(\hat{b}_{{\bf k}1},\hat{b}_{{\bf k}2},\hat{b}_{{\bf k}3},\hat{b}_{{\bf k}4})^T$. 
Incorporating the influence of the AC exchange magnetic field, the field operator of quasiparticles now reads 
\begin{align}
    \hat{\Psi}({\pmb \rho},t)=&\frac{1}{\sqrt{A}}\sum_{\bf k}\left(\Phi_1({\pmb \rho},t),\Phi_3({\pmb \rho},t),\Phi_3({\pmb \rho},t), \Phi_4({\pmb \rho},t)\right)\hat{\pmb a}_{\bf k}\nonumber\\
    =&\frac{1}{\sqrt A}\sum_{\bf k}U({\bf k})e^{i{\bf k}\cdot{\pmb \rho}}\hat{\pmb b}_{\bf k}(t).
    \label{field_operator_scattering}
\end{align}

Figure~\ref{scattering_process} illustrates a one typical scattering process of $\{\hat{a}_{{\bf q}1},\hat{a}_{{\bf q}2},\hat{a}_{{\bf q}3},\hat{a}_{{\bf q}4}\}$ to $\hat{b}_{{\bf k}1}$, including both the intraband and interband processes with a flip of the spin of quasiparticles. We note again that the spin of quasiparticles is no longer $\hbar/2$ but is strongly wave-vector dependent, indicated by the color map on the particle-like and hole-like bands in Fig.~\ref{scattering_process}. 
Since there are no other subsystems to which the spins can go in our
model, and in the absence of the spin-orbit interaction, the net spin of quasiparticles and Cooper pairs should be conserved. Thereby, the spin-flip of quasiparticles can inject the spin carried by the Cooper pairs via the spin torque \eqref{spin_torque_density_operator}. 
The efficiencies of different processes depend on the source frequency. For example, interband transitions across the superconducting gap are inefficient when the frequency $\omega$ of the exchange field is smaller than twice the superconducting gap, which is the typical case for the ferromagnetic resonance (FMR) of ferromagnets.

Substitution of the field operator of quasiparticle \eqref{field_operator_scattering} into the definition of spin-current density  \eqref{spin_current_operator}
yields the equilibrium, AC, and DC spin-current densities pumped by the AC exchange field. We derive these currents in detail in Appendix~\ref{spin_current_appendix}. 
Particularly, the DC spin-current density reads
\begin{align}
&{\bf J}^{\rm DC}_{s}({\pmb \rho})    
=\frac{\hbar^2}{4mA}\sum_{{\bf k}{\bf k}^\prime{\bf q}}\sum_{\xi=\pm}\sum^2_{\alpha\beta\gamma=1}\left({\bf Q}_{\beta\alpha}({\bf k},{{\bf  k}^\prime})\otimes{\bf k}^\prime\right)\nonumber\\
&\times\mathcal{G}^{-\xi}_{\alpha\gamma}({\bf k}^\prime,{\bf q})\mathcal{G}^{\xi}_{\gamma\beta}({\bf q},{\bf k})e^{i({\bf k}^\prime-{\bf k})\cdot{\pmb \rho}}\nonumber\\ 
&\times\frac{f(E_{\beta}(k))-f(E_{\gamma}(q))}{\left(E_{\gamma} (q)-E_{\alpha}( k^\prime)-\xi\hbar\omega+i\delta\right)\left(E_{\beta}( k)-E_{\gamma}( q)+\xi\hbar\omega+i\delta\right)}\nonumber\\
&+{\rm H.c.},
\label{DC_spin_current}
\end{align}
in which the matrix elements
${\bf Q}_{\alpha\beta}({\bf k},{\bf k}^\prime)=\phi_\alpha^\dagger({\bf k})\tau_3{\pmb {\cal S}}\phi_\beta({\bf k}^\prime)$. It is contributed by the magnon absorption and then emission process, and vice versa. When the magnon frequency $\omega<2\Delta_p/\hbar$, the optical transition between the particle bands $E_{1,2}(k)$ and hole bands $E_{3,4}(k)$ is inefficient and is thereby safely disregarded. In this situation, such a DC spin-current density is solely carried by quasiparticles.

On the other hand, the spin-torque density is found by substituting the field operator of quasiparticles \eqref{field_operator_scattering} into Eq.~\eqref{spin_torque_density_operator} and taking the ensemble average, which is detailed in Appendix~\ref{spin_torque_appendix}.
The DC component of the spin-torque density driven by the AC field of frequency $\omega<2\Delta_p/\hbar$ reads 
    \begin{align}
   & \mathbf{T}^{\rm DC}_s({\pmb \rho})
   =\frac{1}{2iA}\sum_{{\bf k}{\bf k}^\prime{\bf q}}\sum_{\xi=\pm}\sum^2_{\alpha\beta\gamma=1}\overline{\bf Q}^{p}_{\beta\alpha}({\bf k},{{\bf  k}^\prime})\nonumber\\
   &\times\mathcal{G}^{-\xi}_{\alpha\gamma}({\bf k}^\prime,{\bf q})\mathcal{G}^{\xi}_{\gamma\beta}({\bf q},{\bf k})e^{i({\bf k}^\prime-{\bf k})\cdot{\pmb \rho}}\nonumber\\
&\times\frac{f(E_{\beta}(k))-f(E_{\gamma}(q))}{\left(E_{\gamma}( q)-E_{\alpha}( k^\prime)-\xi\hbar\omega+i\delta\right)\left(E_{\beta}( k)-E_{\gamma}( q)+\xi\hbar\omega+i\delta\right)}\nonumber\\
&+{\rm H.c.},  
\label{DC_spin_torque}
\end{align}
in which 
\begin{align}
  & \overline{\bf Q}^{p}_{\beta\alpha}({\bf k},{{\bf  k}^\prime})=\phi^\dagger_\beta({\bf k})\Delta_p\left(\begin{array}{cc}
      {\pmb\sigma}   &  \\
         & -{\pmb\sigma^\ast}
    \end{array}\right)\nonumber\\
    &\times\left(\begin{array}{cccc}
         0&0&{e^{i\theta_{{\bf k}^\prime}}}&0  \\
         0&0&0&-e^{-i\theta_{{\bf k}^\prime}}\\
         e^{-i\theta_{{\bf k}^\prime}}&0&0&0\\
         0&-e^{i\theta_{{\bf k}^\prime}}&0&0
    \end{array}\right)\phi_\alpha({\bf k}^\prime).
\end{align}

Both ${\bf J}^{\rm DC}_{s}({\pmb \rho})$ and $\mathbf{T}_s^{\rm DC}({\pmb \rho})$ vanish in the absence of quasiparticles at zero temperature, see the demonstration in Appendix~\ref{spin_current_appendix} and \ref{spin_torque_appendix}. This indicates that both the spin-current density and spin-torque density are related to the population of quasiparticles. This is reasonable since we assume the frequency $\omega< 2\Delta_p/\hbar$ of the AC field is sufficiently low such that no efficient transition between the hole-like bands $E_{3,4}(k)$ and particle-like bands $E_{1,2}(k)$ across the superconducting gap is triggered. In this situation, the spin-flip of quasiparticles among the particle-like bands injects the spin of Cooper pairs that flow away from the source.

\section{Spin pumping of spin supercurrent}
\label{spin_pumping}

Spin pumping refers to the generation of a pure spin current ${\bf J}_s\propto d{\bf M}/dt\times {\bf M}$ of electrons driven by the coherent magnetization ${\bf M}$ dynamics via the interfacial exchange interaction between the magnetic moments and electron spins~\cite{spin_pumping_1}. Here, we propose a theory of pumping of spin supercurrents driven by magnetization dynamics.

Substitution of the coupling constant \eqref{coupling_constants_ex} due to the exchange interaction into Eqs.~\eqref{DC_spin_current} and \eqref{DC_spin_torque}, we obtain, respectively, the DC spin-current and DC spin-torque densities pumped by the coherent magnetization dynamics.

Several features of spin pumping can be inferred from the coupling constants without resorting to explicit calculations. 
The exchange magnetic field in the coupling constant $\sum_{\gamma=1}^2{\cal G}^{-\xi}_{\alpha\gamma}{\cal G}^{\xi}_{\gamma\beta}$ contributes 
\begin{widetext}
\begin{align}
& {\bf H}_{\rm ex}^{-\xi}({\bf k}^\prime-{\bf q})\cdot\pmb{\cal S}\left[\sum_{\gamma=1}^2\phi_\gamma({\bf q})\phi_\gamma^\dagger({\bf q})\right]{\bf H}_{\rm ex}^\xi({\bf q}-{\bf k})\cdot\pmb{\cal S}=\nonumber\\ &\left(\begin{array}{ccc}
{\cal{A}}_{q} \left[({\bf H}_{\rm ex}^{-\xi}\cdot{\bf H}_{\rm ex}^{\xi}){\cal I}+i({\bf H}_{\rm ex}^{-\xi}\times{\bf H}_{\rm ex}^{\xi})\cdot{\pmb\sigma}\right]&{\cal{B}}_q \left( \begin{array}{cc}
{\cal I}&{\pmb\sigma} 
\end{array} \right)\\
&\times\left[\frac{q_y}{\abs{q}}{\cal M}\left(\begin{array}{c} -i{\bf H}_{\rm ex}^{-\xi}\cdot{\bf\tilde H}_{\rm ex}^{\xi}  \\
      {\bf H}_{\rm ex}^{-\xi}\cross{\bf\tilde H}_{\rm ex}^\xi
\end{array}\right)+\frac{q_z}{\abs{q}}\left(\begin{array}{c} -i{\bf H}_{\rm ex}^{-\xi}\cdot{\bf\bar H}_{\rm ex}^{\xi}  \\
      {\bf H}_{\rm ex}^{-\xi}\cross{\bf\bar H}_{\rm ex}^\xi
\end{array}\right)\right] \\
\\
{\cal{B}}_q \left( \begin{array}{cc}
{\cal I}&{\pmb\sigma}^* 
\end{array} \right)\\
\times\left[\frac{q_y}{\abs{q}}{\cal M}^*\left(\begin{array}{c} i{\bf H}_{\rm ex}^{-\xi}\cdot{\bf\tilde H}_{\rm ex}^{\xi}  \\
      {\bf H}_{\rm ex}^{-\xi}\cross{\bf\tilde H}_{\rm ex}^\xi
\end{array}\right)+\frac{q_z}{\abs{q}}\left(\begin{array}{c} i{\bf H}_{\rm ex}^{-\xi}\cdot{\bf\bar H}_{\rm ex}^{\xi}  \\
      {\bf H}_{\rm ex}^{-\xi}\cross{\bf\bar H}_{\rm ex}^\xi
\end{array}\right)\right]& {\cal{C}}_{q} \left[({\bf H}_{\rm ex}^{-\xi}\cdot{\bf H}_{\rm ex}^{\xi}){\cal I}-i({\bf H}_{\rm ex}^{-\xi}\times{\bf H}_{\rm ex}^{\xi})\cdot{\pmb\sigma}^*)\right]
    \end{array}\right).
    \label{HH}
 \end{align}
\end{widetext}
Here, $\mathbf{\tilde H}_{\rm ex}=(-H^x_{\rm ex},H^y_{\rm ex},H^z_{\rm ex})^T$ and $\mathbf{\bar H}_{\rm ex}=(H^x_{\rm ex},-H^y_{\rm ex},H^z_{\rm ex})^T$. We express ${\bf H}_{\rm ex}^{-\xi}={\bf H}_{\rm ex}^{-\xi}({\bf k}^\prime-{\bf q})$ and ${\bf H}_{\rm ex}^{\xi}={\bf H}_{\rm ex}^{\xi}({\bf q}-{\bf k})$ for brevity.    
The coefficients 
\begin{align}
&{\cal A}_q=(E_q+\xi_q)/(2E_q),\nonumber\\
&{\cal B}_q={\Delta_p}/(2E_q),\nonumber\\
&{\cal C}_q={\Delta^2_p}/[2E_q(E_q+\xi_q)],\nonumber
\end{align}
are real, the matrix ${\cal M}={\rm adiag}(-i,1,-1,-i)$, and $\cal I$ is the  $2\times 2$ unit matrix. We then expect combinations such as ${\bf H}_{\rm ex}^{-\xi}\cdot{\bf H}_{\rm ex}^\xi$, ${\bf H}_{\rm ex}^{-\xi}\cdot{\bf\tilde H}_{\rm ex}^\xi$, ${\bf H}_{\rm ex}^{-\xi}\cdot{\bf\bar H}_{\rm ex}^\xi$, ${\bf H}_{\rm ex}^{-\xi}\times{\bf H}_{\rm ex}^\xi$, ${\bf H}_{\rm ex}^{-\xi}\times{\bf\tilde H}_{\rm ex}^\xi$ and ${\bf H}_{\rm ex}^{-\xi}\times{\bf \bar H}_{\rm ex}^\xi$ contribute to the DC spin-current and spin-torque densities. When $\Delta_p\rightarrow 0$, ${\cal A}_q\rightarrow 1$, ${\cal B}_q\rightarrow 0$, and ${\cal C}_q\rightarrow 0$, such that only the terms $({\bf H}_{\rm ex}^{-\xi}\cdot{\bf H}_{\rm ex}^{\xi}){\cal I}+i({\bf H}_{\rm ex}^{-\xi}\times{\bf H}_{\rm ex}^{\xi})\cdot{\pmb\sigma}$ survives in the coupling constants \eqref{HH}. Nevertheless, only $i({\bf H}_{\rm ex}^{-\xi}\times{\bf H}_{\rm ex}^{\xi})\cdot{\pmb\sigma}$ contributes to the generation of spin current by further considering the factor ${\bf Q}_{\alpha\beta}({\bf k},{\bf k}')$ in the DC spin current \eqref{DC_spin_current}, which is consistent with the spin pumping into the normal metal~\cite{spin_pumping_1}. The existence of the triplet order parameter then significantly alters the features of conventional spin pumping~\cite{spin_pumping_1} by introducing many new contributions.

For the prototypical unitary $p$-wave superconductor, we consider in the Hamiltonian \eqref{Hamiltonian}, all the possible contributions of the coherent dynamics of the magnetization to the pumped spin-current and spin-torque densities are summarized in Table~\ref{tab:selection-rules_exchange}. The polarizations to the pumped spin density and spin torque strongly depend on the combinations of $\mathbf{M}^*$ with $\mathbf{M}=(M_x,M_y,M_z)^T$, $\mathbf{\tilde M}=(-M_x,M_y,M_z)^T$, and $\mathbf{\bar M}=(M_x,-M_y,M_z)^T$. These contributions can act as a ``selection rule" that guides the analysis of spin pumping to the triplet superconductor (or superconductors with the triplet Cooper pairing)~\cite{KRjeon}.

\begin{table}[htp!]
\centering
\caption{Possible contributions of the coherent magnetization dynamics to the DC spin-current and spin-torque densities in the prototypical unitary $p$-wave superconductor. The superscript on the spin-current/torque indicates the spin-polarization direction. In the table, $\mathbf{M}=(M_x,M_y,M_z)^T$, $\mathbf{\tilde M}=(-M_x,M_y,M_z)^T$, and $\mathbf{\bar M}=(M_x,-M_y,M_z)^T$.}
\label{tab:selection-rules_exchange}
\begin{tabular}{@{}ll@{}}
\toprule
~\textbf{spin current/torque} &~~~~~~~~\textbf{governed by}\\
\midrule
~${\bf J}_s^z$,~~~ ${\bf T}_s^z$ &~~~${\bf M}^*{\cdot}{\bf M}$, ~${\bf M}^*{\cdot}\tilde{\bf M}$, ~${\bf M}^*{\cdot}\bar{\bf M}$\\
~${\bf J}_s^{x,y}$,~~${\bf T}_s^{x,y}$ &~~~${\bf M}^*{\times}{\bf M}$, ~${\bf M}^*{\times}\tilde{\bf M}$,  ~${\bf M}^*{\times}\bar{\bf M}$ \\
&~~~~($x/y$ component)\\
~${\bf J}_s^{z}$,~~~ ${\bf T}_s^{z}$ &~~~${\bf M}^*{\times}{\bf M}$, ~${\bf M}^*{\times}\tilde{\bf M}$, ~${\bf M}^*{\times}\bar{\bf M}$\\
&~~~~($z$ component)\\
\bottomrule
\end{tabular}
\end{table}

\begin{figure}[htp!]
\centering
\includegraphics[width=1.0\linewidth,clip, trim=0cm 4cm 0cm 2.5cm]{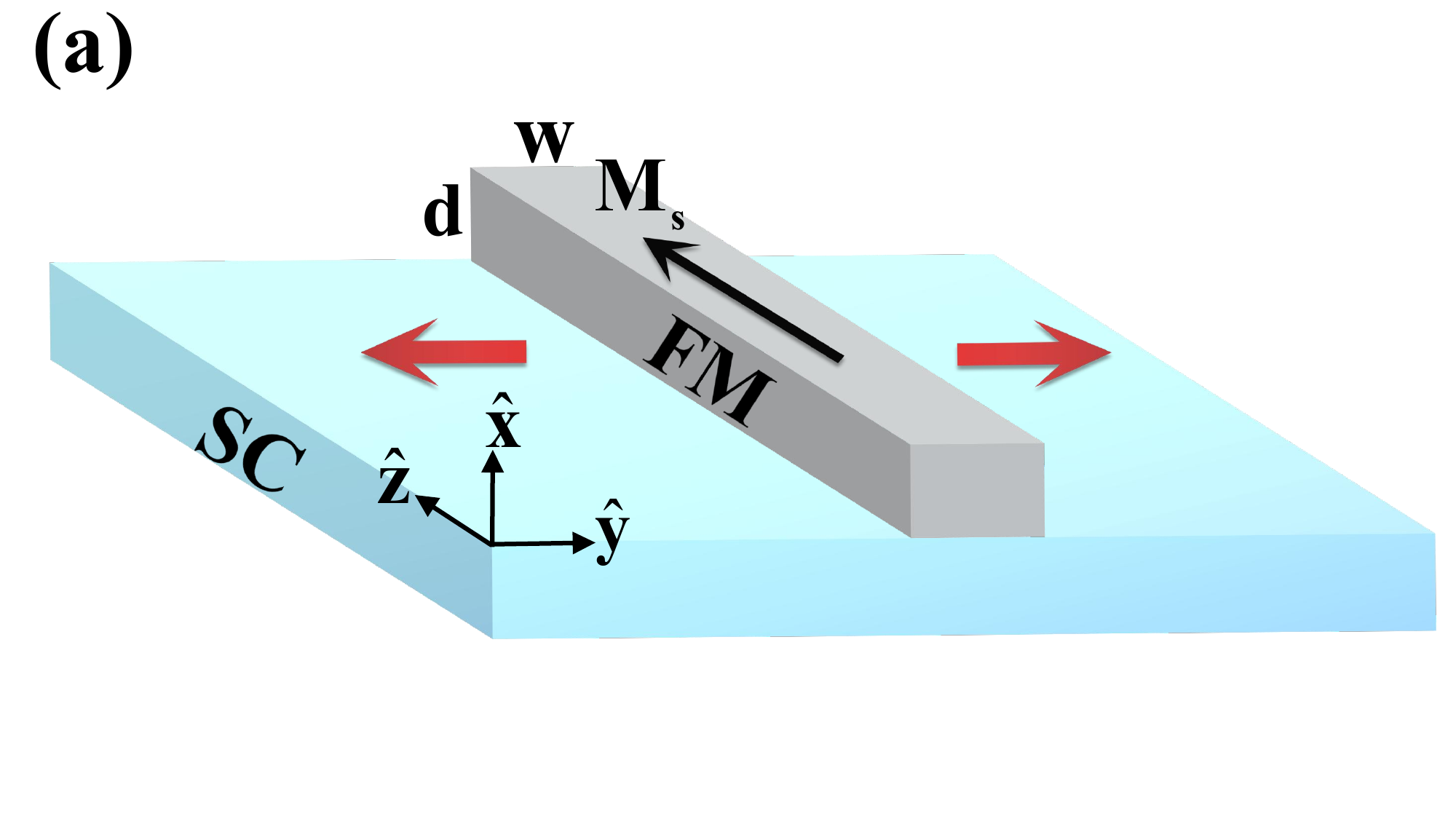}
\caption{Schematic of pumping the spin current carried by the quasiparticles and Cooper pairs in a unitary $p$-wave superconductor, driven by the coherent magnetization dynamics of the ferromagnetic resonance of the magnetic nanowire via an interfacial exchange field. The width and thickness of the wire are $d$ and $w$, respectively. The saturation magnetization $M_s$, denoted by the thick black arrow, is biased to the wire $\pm \hat{\bf z}$-direction. The thick red arrow represents the longitudinal spin pumping that contains both the dissipative and dissipationless spin current.} 
\label{exchange_field}
\end{figure}

We now apply the general analysis Table~\ref{tab:selection-rules_exchange} to a typical configuration of spin pumping by specifying a configuration in Fig.~\ref{exchange_field} to illustrate the principle of pumping non-dissipative spin supercurrent, in which a ferromagnetic nanowire of thickness $d$ and width $w$ with the saturation magnetization $M_s\hat{\bf z}$ along the wire $\hat{\bf z}$-direction is fabricated on top of the $p$-wave triplet superconductor.  The coherent magnetization $\mathbf{M}({\bf r},t)={M}_{\perp}(-i/\xi^2_m,1,0)^{\rm T}e^{-i\omega t} +{\rm H.c.}$ precesses around the saturation magnetization $M_s\hat{\bf z}$ with the FMR frequency $\omega$, amplitude of the transverse magnetization $M_{\perp}$, and the ellipticity $\xi^2_m$ that depends on the geometry of the ferromagnetic nanostructures.
For the nanowire, the ellipticity $\xi^2_m=1$ when $w=d$. The exchange magnetic field is then confined to  the region  $-w/2<y<w/2$ and has the Fourier components 
\[
{\bf H}^\xi_{\rm ex}({\bf k})=J_{\rm ex}\hbar\pi  {M}_\perp\frac{\sin(k_yw/2)}{k_y}\left(-i\frac{\xi}{\xi^2_m},1,0\right)^T\delta(k_z).
\]

For such a one-dimensional exchange magnetic field, the $z$-component of ${\bf M}^*\times {\bf M}$ exists, which drives the spin-current and spin-torque with the spin polarization along the saturation magnetization $\hat{\bf z}$-direction. Further analysis shows that they flow along the longitudinal $\hat{\bf y}$-direction. ${\bf M}^*\cdot{\bf M}$ also exists, noting  ${\bf M}^*\cdot\tilde{\bf M}=0$ and  ${\bf M}^*\cdot\bar{\bf M}=0$ with $\mathbf{M}^{\xi}({\bf r},t)={M}_{\perp}(-i/\xi^2_m,1,0)^{\rm T}$, which also contributes to the spin-current and spin-torque densities polarized along the $\hat{\bf z}$-direction, but the flow direction is transverse, i.e., along the magnetic wire $\hat{\bf z}$-direction according to Table~\ref{tab:selection-rules_exchange}.

We substantiate such expectation and illustrate our results using the parameters of a typical $p$-wave triplet superconductor candidate, \textit{i.e.},  LaAlO$_3$/KTaO$_3$ superconducting interface~\cite{C_Liu,Z_Chen,interface,Hanwei}. According to the experiment~\cite{interface}, the typical electron density $n_e\approx1.8\times 10^{14}~ \rm cm^{-2}$ of the two-dimensional electron gas corresponds to the Fermi energy $E_F\approx431~\rm meV$.
We model the temperature dependence of  the $p$-wave triplet order parameter according to the experimental measurements~\cite{interface}, which is well fitted by a generalized BCS model with ${T_c}=2.05~{\rm K}$ and $2\Delta_p(T=0)/(k_bT_c)\approx 4.3$, i.e., 
\[
\Delta_p(T)=\Delta_p(T=0)\tanh(1.74\sqrt{T_c/T-1}).
\]

\begin{figure}[htp!]
    \centering
    \begin{minipage}[c]{0.492\columnwidth}
        \centering
        \includegraphics[width=\linewidth, clip, trim=0.2cm 0cm 1cm 0cm]{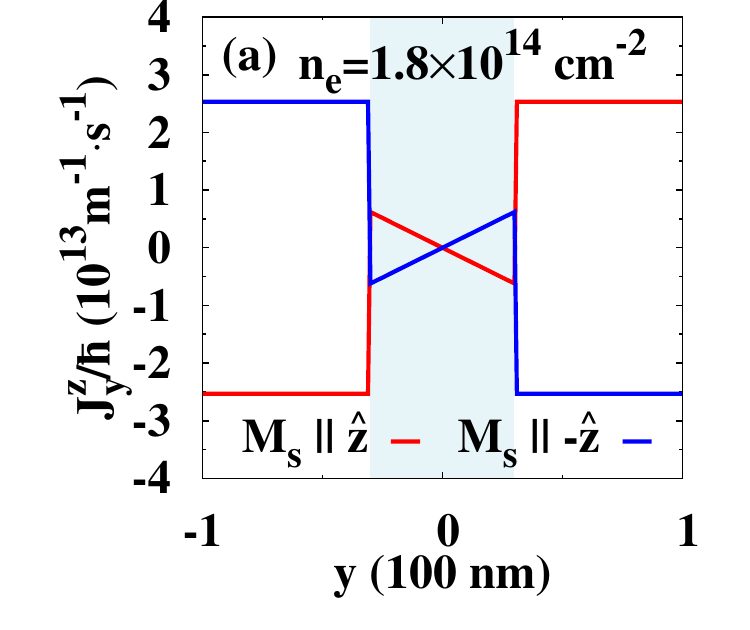}
    \end{minipage}
    \begin{minipage}[c]{0.492\columnwidth}
        \centering
        \includegraphics[width=\linewidth, clip, trim=0.2cm 0cm 1cm 0cm]{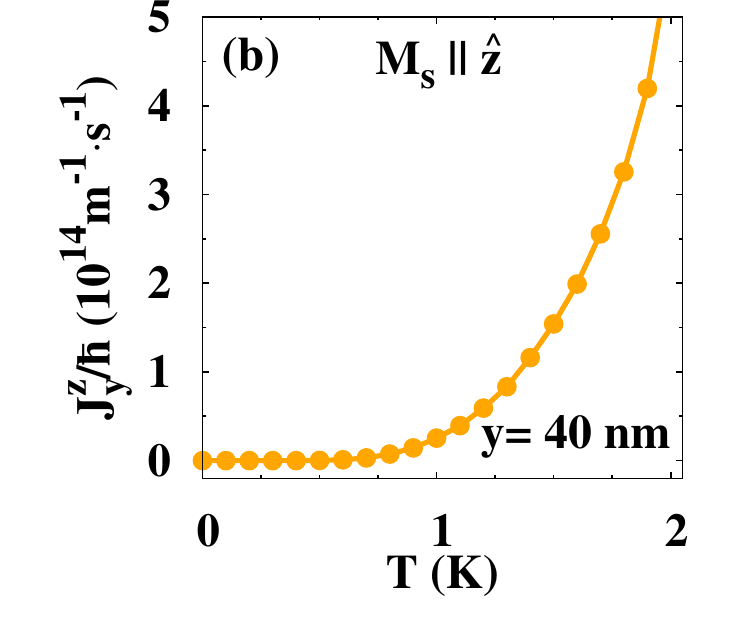}
    \end{minipage}
    \begin{minipage}[b]{0.492\columnwidth}
        \centering
        \includegraphics[width=\linewidth, clip, trim=0.2cm 0cm 1cm 0cm]{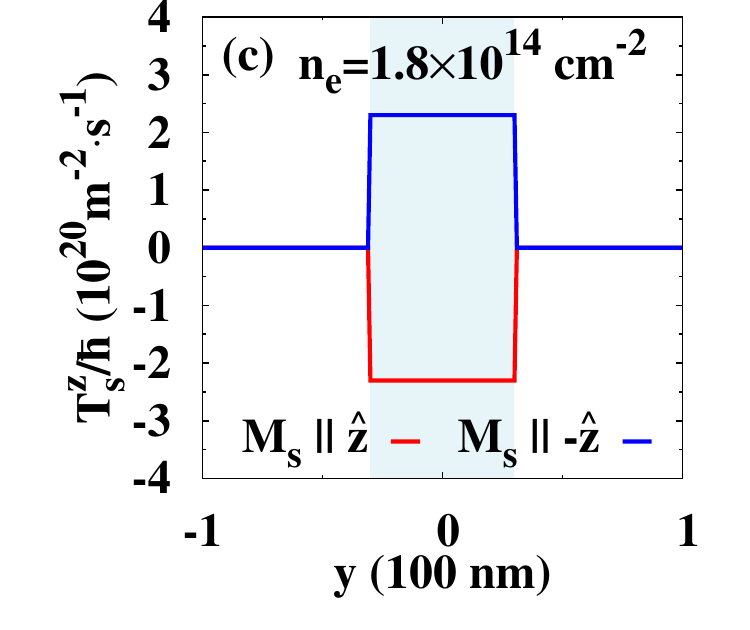}
    \end{minipage}
    \begin{minipage}[b]{0.492\columnwidth}
        \centering
        \includegraphics[width=\linewidth, clip, trim=0.2cm 0cm 1cm 0cm]{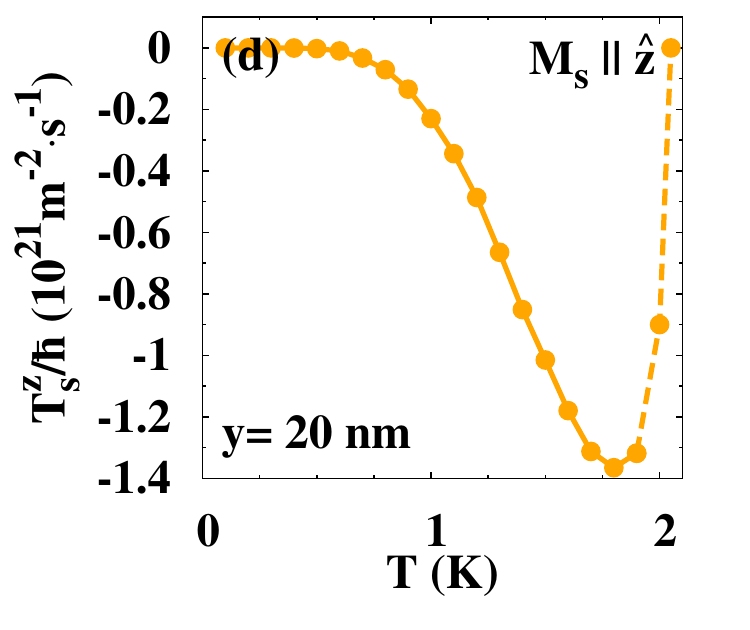}
    \end{minipage}
    \begin{minipage}[b]{0.492\columnwidth}
        \centering
        \includegraphics[width=\linewidth, clip, trim=0.2cm 0cm 1cm 0cm]{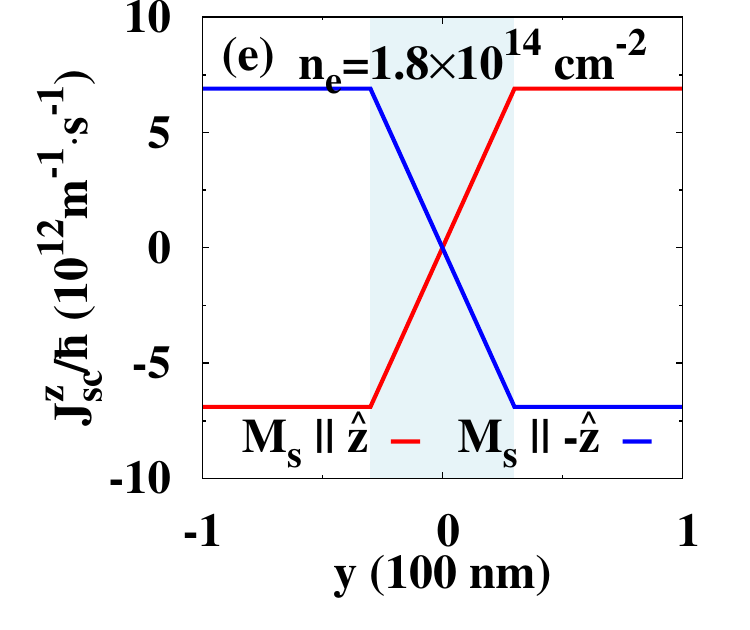}
    \end{minipage}
    \begin{minipage}[b]{0.492\columnwidth}
        \centering
        \includegraphics[width=\linewidth, clip, trim=0.2cm 0cm 1cm 0cm]{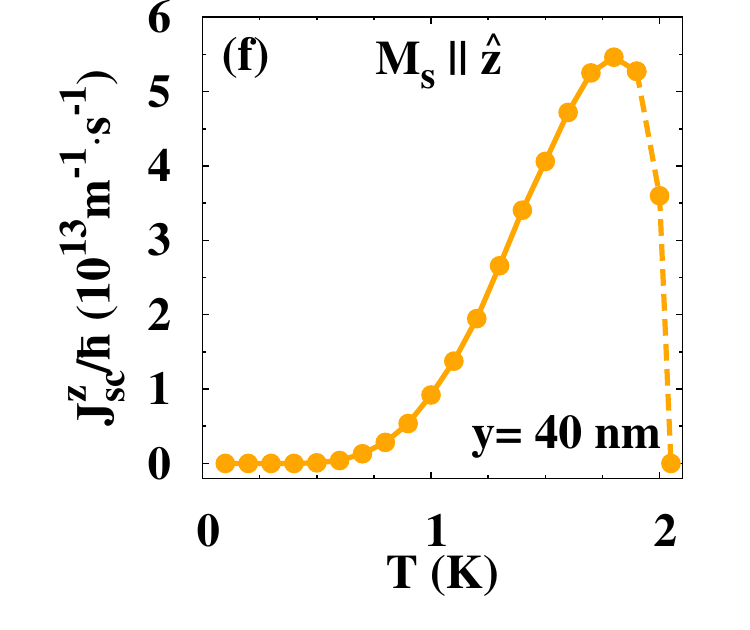}
    \end{minipage}
     \includegraphics[width=0.95\linewidth, clip, trim=0cm 2.8cm 0cm 2.8cm]{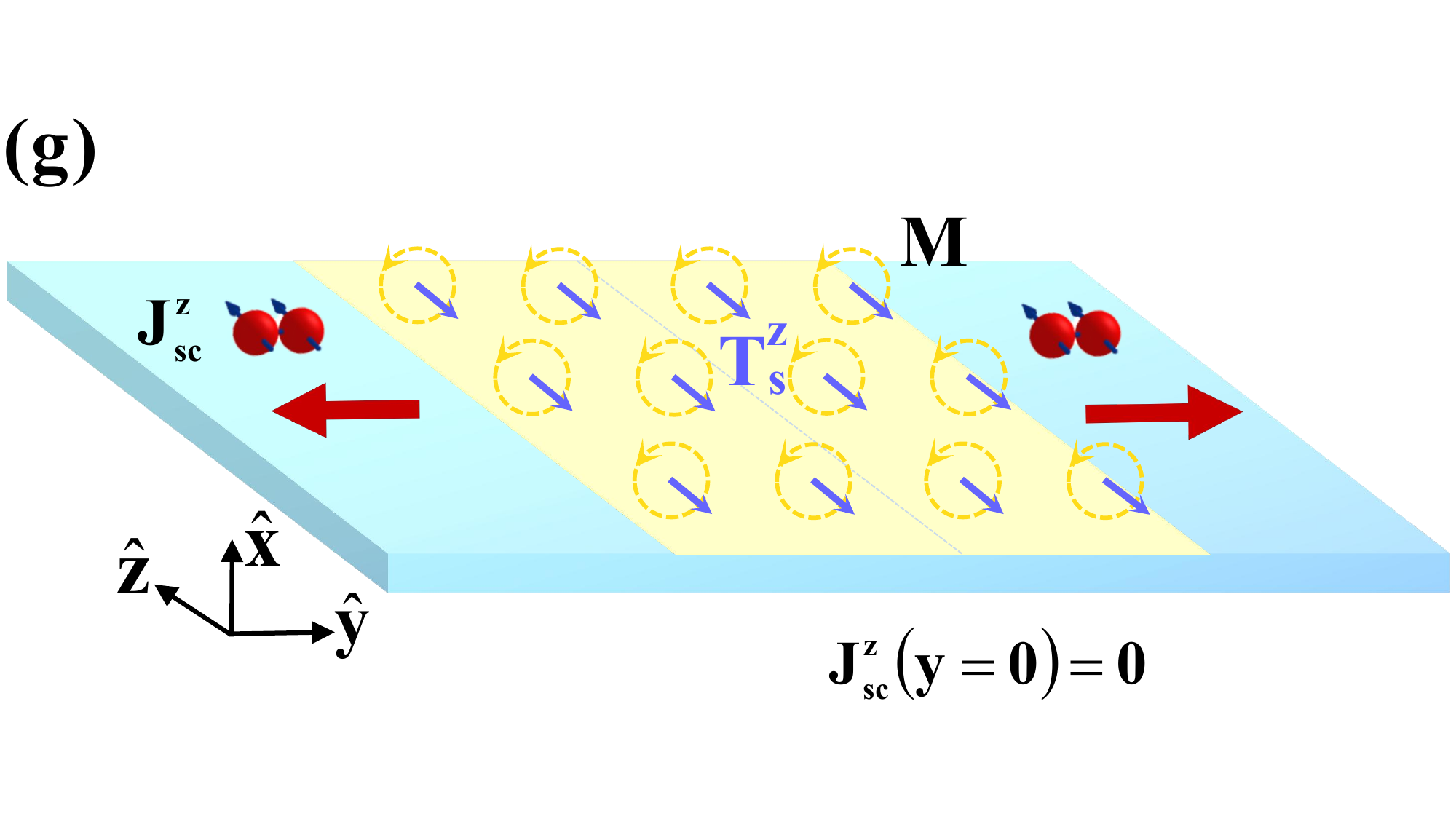}
    \caption{Pumping of spin current and spin supercurrent into the $p$-wave triplet superconductors, driven by the dynamical interfacial exchange field due to the ferromagnetic resonance of magnetic nanowires. The shadow region in (a), (c), and (e) indicates the region of the dynamical interfacial exchange field covered by the magnetic nanowire. (a) and (b) plot the spatial and temperature dependencies of the pumped spin current that is polarized along the saturation magnetization $\pm\hat{\bf z}$-direction. (a) is calculated at the temperature ${ T} =1~{\rm K}$, noting the superconducting transition temperature $T_c=2.05$~K. (c) addresses the spatial profile of the spin-torque density when $T=1~{\rm K}$, which is also polarized along the saturation magnetization $\hat{\bf z}$-direction. The temperature dependence of the spin-torque density inside the pumping region is shown in (d). (e) illustrates the spatial dependence of the spin supercurrent by expressing the spin-torque density ${\bf T}_s(y)=-\partial_y{\bf J}_{sc}(y)$ when $T = 1~{\rm K}$, under the assumption ${\bf J}_{sc}(y=0)=0$.  (f) illustrates the temperature dependence of the spin supercurrent close to the pumping region. (g) illustrates the spin torque and pumped spin supercurrent when ${\bf M}_s\parallel \hat{\bf z}$. The material parameters used for the calculation are given in the text.}
    \label{Magnetic_Field}
\end{figure}

Figure~\ref{Magnetic_Field} illustrates the spin-current and spin-torque densities pumped by the AC exchange field of frequency $\omega= 2\pi\times 28~{\rm GHz}$, emitted by the FMR by a small transverse magnetization $M_\perp=0.1M_s$ to the saturation magnetization $M_s$ of the CoFeB magnetic nanowire of geometries $d=w=60~{\rm nm}$,  which is biased by a static magnetic field $\mu_0 H_0=0.1~{\rm T}$ applied along the wire $\pm\hat{\bf z}$-direction. The field frequency corresponds to $\hbar\omega\approx0.1~{\rm meV}$, which is smaller than $2\Delta_p>0.5$~meV in our calculation. 
The exchange coupling constant $J_{\rm ex}=2.4\times 10^5\ \rm{m/ (s\cdot A)}$  is estimated from  the exchange splitting energy $\Delta=0.2~{\rm meV}$ that is realized in the experiments~\cite{splitting_energy1,splitting_energy2}. The saturation magnetization of CoFeB $\mu_0 M_s= 1.6~{\rm T} $~\cite{CoFeB_1,CoFeB_2,CoFeB_3}.

As illustrated in Fig.~\ref{Magnetic_Field}(a),  when ${\bf M}_s\parallel\hat{\bf z}$, the pumped spin current polarized along the saturation magnetization $\hat{\bf z}$-direction flows outward from the pumped region with equal magnitude and opposite flow directions at both sides. 
Reversing the saturation magnetization $({\bf M_s}\parallel -\hat{\bf z})$ flips the spin-polarization direction accordingly since the spin direction of magnons in the ferromagnetic resonance is reversed. Furthermore, as shown in Fig.~\ref{Magnetic_Field}(b), the temperature affects the efficiency of generating spin current strongly: on one hand, the spin-current density is suppressed to vanish when $T\rightarrow 0$; on the other hand,  the magnitude of the spin current increases significantly with the increase of the temperature. This implies that the spin current defined in Eq.~\eqref{DC_spin_current} is carried dominantly by the quasiparticles in our regime with $\hbar \omega<2\Delta_p$, which are thermally excited when $T\rightarrow T_c$, while they are much suppressed when $T\ll T_c$. These features agree with the spin-pumping experiments in conventional superconductors~\cite{probe_spin}, in which the spin current is carried by quasiparticles when below the superconducting transition temperature. 

Similar to the spin polarization of the injected quasiparticles, the injected spin-torque density is polarized along the saturation magnetization $\hat{\bf z}\mbox{-}$direction, which is reversed when the direction of the saturation magnetization is reversed, as illustrated in Fig.~\ref{Magnetic_Field}(c).
However, different from the spatial distribution of the spin-current density in Fig.~\ref{Magnetic_Field}(a), the spin-torque density is symmetric with respect to the center of the pumping region, which remains nearly constant inside the pumping region, but drops to a much smaller magnitude outside this region. As in Fig.~\ref{Magnetic_Field}(d), when $T\rightarrow 0$, the spin-torque density vanishes, similar to the spin-current density carried by quasiparticles [Fig.~\ref{Magnetic_Field}(b)].  This implies that the spins of Cooper pairs are injected during the scattering of quasiparticles, which makes an analogy to the Andreev reflection. With the increase of the temperature, the spin-torque density first increases due to the increased thermal excitation of quasiparticles, as shown in Fig.~\ref{Magnetic_Field}(d), but decreases when $T\rightarrow T_c$, since the spin torque density is proportional to the triplet order parameter $ {\bf T}^z_s({\pmb \rho})\propto\Delta_p$ as in Eq.~\eqref{spin_torque_density_operator}. This behavior differs markedly from that of the spin-current density, which increases monotonically with increasing temperature [Fig.~\ref{Magnetic_Field}(b)] when below $T_c$.

We can now make an analogy to the Andreev reflection~\cite{Tinkham_book,BTK,Andreev_reflection} by interpreting the spin-torque density ${\bf T}_s({\pmb \rho})$ as the divergence  of a spin supercurrent density ${\bf J}_{sc}({\pmb \rho})$, i.e,   
 \begin{align}
 {\bf T}_s({\pmb \rho})=-{\pmb\nabla}\cdot {\bf J}_{sc}({\pmb \rho}),
 \end{align}
such that in the spin continuity equation \eqref{spin_continuity_equation}, the spin becomes conserved by the definition of the total spin-current density ${\bf J}_s({\pmb \rho})+{\bf J}_{sc}({\pmb \rho})$.

The spin supercurrent density ${\bf J}_{sc}({\pmb \rho})$ can be easily calculated in our spin-pumping setup.  According to the spatial profile of the spin-torque density ${\bf T}_s(y)$ as shown in Fig.~\ref{Magnetic_Field}(c),  we calculate the spin supercurrent  ${\bf J}_{sc}(y)={\bf J}_{sc}(y=0)-\int^y_{0}dy^\prime {\bf T}_s(y^\prime)$ when $T=1$~K. Considering the mirror symmetry, 
we assume ${\bf J}_{sc}(y=0)=0$ and  plot the obtained spin supercurrent density in Fig.~\ref{Magnetic_Field}(e). Similar to the quasiparticle spin current ${\bf J}_s(y)$, the pumped spin supercurrent polarized along the saturation magnetization $\hat{\bf z}$-direction flows outward from the pumped region with equal magnitude and opposite flow directions at both sides when ${\bf M}_s\parallel\hat{\bf z}$. Reversing the saturation magnetization $({\bf M}_s\parallel -\hat{\bf z})$ flips the spin-polarization direction accordingly due to the reversed spin direction of FMR magnons. 
Figure~\ref{Magnetic_Field}(f) shows the temperature dependence of the spin supercurrent density ${\bf J}_{sc}$  close to the pumping region, which vanishes when $T\rightarrow 0$ but exhibits a unique maximum when $T\lesssim T_c$.
When $T\rightarrow T_c$, the spin supercurrent density vanishes as well as the spin torque density, since the spin supercurrent density is proportional to the triplet superconducting order parameter.

The generation of spin supercurrent extracts additional energy from the coherent magnetization dynamics, thereby enhancing the Gilbert damping~\cite{spin_pumping_1,KRjeon}. Figure~\ref{Magnetic_Field}(g) illustrates and summarizes all the features of the generated spin torque and pumped spin supercurrent by coherent magnetization dynamics when ${\bf M}_s\parallel \hat{\bf z}$.

\section{Discussion and conclusion}
\label{discussion_conclusion}

In conclusion, we develop a general theory of spin pumping of DC \textit{spin supercurrent} by coherent magnetization dynamics in prototypical triplet superconductors. We find many features that go beyond the conventional spin pumping that focus on the spin current carried by normal electrons or quasiparticles~\cite{spin_pumping_1}. In the existence of the triplet superconducting order parameter, the spin pumping does not simply transfer the FMR magnon spin $d{\bf M}/dt\times {\bf M}$ into the Cooper pairs, but all the other combinations of magnetization components are allowed (Table~\ref{tab:selection-rules_exchange}), thereby going beyond the conventional spin pumping~\cite{spin_pumping_1}. Our theory is based on the robust conservation law, in which we find that the triplet superconducting order parameter provides a sink to the spin density carried by quasiparticles, which shares exactly the same structure as the charge dynamics in a conventional superconductor, in which the singlet superconducting order parameter provides a sink to the charge density carried by quasiparticles. The latter effect has been exploited to understand the formation of a charge supercurrent during Andreev reflection between a normal metal and conventional superconductors. Our formalism thereby establishes a solid basis for analogous spin dynamics in triplet superconductors. It is also crucial that our proposed effects work in the unitary triplet superconductors, in which the Cooper pairs carry no spin polarization at equilibrium, a system that is not easy to excite spin supercurrent by temperature gradient or charge supercurrent. The proposed spin pumping of spin supercurrent holds a high efficiency: The spin current is of the same order as the spin Hall current generated by an electric field of $1~{\rm kV/cm}$ and a spin Hall conductivity $\sigma_x^y=10^7~(\Omega\cdot {\rm m})^{-1}$, which should be easily measurable~\cite{efficiency1,efficiency2,efficiency3}. Our results open new pathways to
rapidly and energy-efficiently generate and control spin supercurrent in intrinsic triplet superconductors  
and related triplet superconductivity in heterostructures.

\begin{acknowledgments}
This work is financially supported by the National Key Research and Development Program of China under Grant No.~2023YFA1406600 and the National Natural Science Foundation of China under Grant No.~12374109. We thank I. V. Bobkova for the inspiring discussion.
\end{acknowledgments}

\begin{appendix}

\section{Spin-current density}

\label{spin_current_appendix}

Substitution of the field operator \eqref{field_operator_scattering} into the definition of spin current \eqref{spin_current_operator} and taking the ensemble average $\langle \cdots \rangle$
yields the equilibrium, AC, and DC spin currents pumped by the AC exchange field.

To calculate the linear and nonlinear responses, we express the operator $\hat{\bf b}_{\bf k}=\hat{\bf a}_{\bf k}+\hat{\bf b}_{\bf k}^{(1)}+\hat{\bf b}_{\bf k}^{(2)}+\cdots$, in which $ \hat{\bf b}_{\bf k}^{(1)}$ and $ \hat{\bf b}_{\bf k}^{(2)}$ represent the contributions from the first- and second-order scattering processes.

\subsection{Spin-current density at the equilibrium}

With the field operator of quasiparticles \eqref{field_operator_scattering}, the spin-current density  at the equilibrium reads 
\begin{align}
   {\bf J}_{s}^{(0)}({\pmb \rho})&=\frac{i\hbar^2}{8mA}\sum_{{\bf k}}\left\langle \hat{\textbf{\emph a}}^\dagger_{\bf k} U^\dagger_{\bf k}({\pmb \rho},t)(\overleftarrow{\pmb\nabla}-\overrightarrow{\pmb{\nabla}})\tau_3\pmb{\cal S}
      U_{\bf k}({\pmb \rho},t)\hat{\textbf{\emph a}}_{\bf k}\right\rangle\nonumber\\
&=\frac{\hbar^2}{8mA}\sum_{{\bf k}}\sum^4_{\alpha=1}f(E_{\alpha}(k)){\bf Q}_{\alpha\alpha}({\bf k},{\bf k})\otimes{\bf k}+{\rm H.c.},
\end{align}
in which the matrix elements
${\bf Q}_{\alpha\beta}({\bf k},{\bf k}^\prime)=\phi_\alpha^\dagger({\bf k})\tau_3\pmb{\cal S}\phi_\beta({\bf k}^\prime)$.
As expected, this spin-current density at the equilibrium vanishes. 
Explicitly, when the spin polarization is along the $\hat{\bf x}$-direction, the spin-current density  
\begin{align}
  {\bf J}^{x,(0)}_{s}&=\frac{\hbar^2}{8mA}\sum_{{\bf k}} (k_y\hat{\bf y}+{k_z\hat{\bf z}})\cos\theta_{\bf  k}\nonumber\\
  &\times\left[\frac{\xi_{ k}}{E_1({k})}\big(1-2f(E_1({ k}))\big)+\frac{\xi_{k}}{E_2({ k})}\big(2f(E_2({ k}))-1\big)\right]\nonumber\\
  &+{\rm H.c.},
\end{align}
in which we use $\langle\hat{a}^\dagger_{{\bf k}1}  \hat{a}_{{\bf k}1}\rangle=f(E_1({k}))$ and $  \langle\hat{a}_{{-\bf k}1}  \hat{a}^\dagger_{{-\bf k}1}\rangle =f(-E_1({ k}))$. 
It vanishes since $E_1({ k})=E_2({k})$.
When the spin polarization is along the $\hat{\bf y}$-direction, the spin-current density 
\begin{align}
  {\bf J}^{y,(0)}_{s}&=\frac{\hbar^2}{8mA}\sum_{{\bf k}} (k_y\hat{\bf y}+{k_z\hat{\bf z}})\sin\theta_{\bf k}\nonumber\\
  &\times\left[\frac{\xi_{ k}}{E_1({ k})}\big(2f(E_1({ k}))-1\big)+\frac{\xi_{k}}{E_2({ k})}\big(1-2f(E_2({ k}))\big)\right]\nonumber\\
  &+{\rm H.c.}   
\end{align}
also vanishes with $E_1({ k})=E_2({k})$. Similarly, the spin-current density ${\bf J}_s^{z,(0)}$ vanishes.

\subsection{Spin-current density in linear-response regime}

In the linear-response regime, the excited spin-current density  by the AC magnetic field according to Eq.~\eqref{spin_current_operator} reads 
\begin{widetext}
\begin{align}
{\bf J}_{s}^{(1)}({\pmb \rho})&=\frac{i\hbar^2}{8mA}\sum_{{\bf k}}\sum_{{\bf k}^\prime}\left\langle\hat{\textbf{\emph a}}^\dagger_{ \bf k}U^\dagger_{\bf k}({\pmb \rho},t)(\overleftarrow{\pmb\nabla}-\overrightarrow{\pmb{\nabla}})\tau_3\pmb{\cal S}U_{{\bf k}^\prime}({\pmb \rho},t) \hat{\textbf{\emph b}}^{(1)}_{\bf k^\prime}\right\rangle+{\rm H.c.}\nonumber\\
&=\frac{\hbar^2}{8mA}\sum_{{\bf k}}\sum_{{\bf k}^\prime}\sum_{\xi=\pm}\sum^4_{\alpha,\beta=1}\left({\bf Q}_{\beta\alpha}({\bf k},{\bf  k}^\prime)\otimes({\bf k}+{\bf  k}^\prime)  \right)\mathcal{G}^\xi_{\alpha\beta}({\bf k}^\prime,{\bf k})\frac{e^{i\left(({\bf k}^\prime-{\bf k})\cdot {\pmb \rho}-\xi\omega t\right)}}{\xi\hbar\omega-E_{\alpha}( k^\prime)+E_{\beta}( k) +i\delta}f(E_{\beta}( k))+{\rm H.c.}\nonumber\\
&\approx\frac{\hbar^2}{8mA}\sum_{{\bf k}}\sum_{{\bf k}^\prime}\sum_{\xi=\pm}\Bigg{[}\sum_{\alpha,\beta=\{1,2\}}\left({\bf Q}_{\beta\alpha}({\bf k},{\bf  k}^\prime)\otimes({\bf k}+{\bf  k}^\prime)  \right)\mathcal{G}^\xi_{\alpha\beta}({\bf k}^\prime,{\bf k})\frac{e^{i\left(({\bf k}^\prime-{\bf k})\cdot {\pmb \rho}-\xi\omega t\right)}}{\xi\hbar\omega-E_{\alpha}( k^\prime)+E_{\beta}( k) +i\delta}f(E_{\beta}( k))\nonumber\\
&+\sum_{\bar\alpha,\bar\beta=\{3,4\}}\left({\bf Q}_{\Bar{\beta}\Bar{\alpha}}({\bf k},{\bf  k}^\prime)\otimes({\bf k}+{\bf  k}^\prime)\right)\mathcal{G}^\xi_{\Bar{\alpha}\Bar{\beta}}({\bf k}^\prime,{\bf k})\frac{e^{i\left(({\bf k}^\prime-{\bf k})\cdot {\pmb \rho}-\xi\omega t\right)}}{\xi\hbar\omega-E_{\Bar{\alpha}}( k^\prime)+E_{\Bar{\beta} }(k) +i\delta}f(E_{\Bar{\beta} }(k))\Bigg{]}+{\rm H.c.},
\label{spin_current_linear_response}
\end{align}
in which, when the magnon frequency $\omega\ll 2\Delta_p/\hbar$, the optical transition between the particle $\alpha=\{1,2\}$ and hole $\bar\alpha=\{3,4\}$ bands is disregarded. 
 Thereby, the contributions to ${\bf J}_s^{(1)}({\pmb \rho})$ arise exclusively from optical transitions among the particles and among the holes. 
 By the particle-hole symmetry,  Eq.~\eqref{spin_current_linear_response} can be calculated only in the particle space by 
 \begin{align}
     {\bf J}_{s}^{(1)}({\pmb \rho})&=\frac{\hbar^2}{8mA}\sum_{{\bf k}{\bf k}^\prime}\sum_{\xi=\pm}\sum_{\alpha,\beta=\{1,2\}}\Big({\bf Q}_{\beta\alpha}({\bf k},{\bf  k}^\prime)\otimes({\bf k}+{\bf  k}^\prime)  \Big)\mathcal{G}^\xi_{\alpha\beta}({\bf k}^\prime,{\bf k})\frac{e^{i\left(({\bf k}^\prime-{\bf k})\cdot {\pmb \rho}-\xi\omega t)\right)}}{\xi\hbar\omega-E_{\alpha}( k^\prime)+E_{\beta}( k) +i\delta}
     \big[2f(E_{\beta}(k))-1\big]+{\rm H.c.}.
\label{spin_current_linear_response_simplfy}
 \end{align}

We now show that ${\bf J}_{s}^{(1)}({\pmb \rho})$ is solely carried by quasiparticles, which is absent when $T\rightarrow 0$ with $f(E_{\beta }(k))=0$ and $f(E_{\bar{\beta} }(k))=1$, i.e., no quasiparticles exist.  Equation~\eqref{spin_current_linear_response}  then becomes
\begin{align}
   {\bf J}_{s}^{(1)}(T\rightarrow 0)
   &=\frac{\hbar^2}{8mA}\sum_{{\bf k}{\bf k}^\prime}\sum_{\bar\alpha\bar\beta}\sum_{\xi=\pm}\left({\bf Q}_{\Bar{\beta}\Bar{\alpha}}({\bf k},{\bf  k}^\prime)\otimes({\bf k}+{\bf  k}^\prime)\right)\mathcal{G}^\xi_{\Bar{\alpha}\Bar{\beta}}({\bf k}^\prime,{\bf k})\frac{e^{i\left(({\bf k}^\prime-{\bf k})\cdot {\pmb \rho}-\xi\omega t)\right)}}{\xi\hbar\omega-E_{\Bar{\alpha}}( k^\prime)+E_{\Bar{\beta} }(k) +i\delta}\nonumber\\
   &+\frac{\hbar^2}{8mA}\sum_{{\bf k}{\bf k}^\prime}\sum_{\bar\alpha\bar\beta}\sum_{\xi=\pm}\left({\bf Q}_{\Bar{\alpha}\Bar{\beta}}({\bf  k}^\prime,{\bf k})\otimes({\bf k}+{\bf  k}^\prime)\right)\mathcal{G}^{-\xi}_{\Bar{\beta}\Bar{\alpha}}({\bf k},{\bf k}^\prime)\frac{e^{i\left(({\bf k}-{\bf k}^\prime)\cdot {\pmb \rho}+\xi\omega t)\right)}}{\xi\hbar\omega-E_{\Bar{\alpha}}( k^\prime)+E_{\Bar{\beta} }(k) -i\delta}\nonumber\\
   &=\frac{\hbar^2}{8mA}\sum_{{\bf k}{\bf k}^\prime}\sum_{\bar\alpha\bar\beta}\sum_{\xi=\pm}\left({\bf Q}_{\Bar{\beta}\Bar{\alpha}}({\bf k},{\bf  k}^\prime)\otimes({\bf k}+{\bf  k}^\prime)\right)\mathcal{G}^\xi_{\Bar{\alpha}\Bar{\beta}}({\bf k}^\prime,{\bf k})\frac{e^{i\left(({\bf k}^\prime-{\bf k})\cdot {\pmb \rho}-\xi\omega t)\right)}}{\xi\hbar\omega-E_{\Bar{\alpha}}( k^\prime)+E_{\Bar{\beta}}( k) +i\delta}\nonumber\\
   &+\frac{\hbar^2}{8mA}\sum_{{\bf k}{\bf k}^\prime}\sum_{\bar\alpha\bar\beta}\sum_{\xi=\pm}\left({\bf Q}_{\Bar{\beta}\Bar{\alpha}}({\bf k},{\bf  k}^\prime)\otimes({\bf k}+{\bf  k}^\prime)\right)\mathcal{G}^\xi_{\Bar{\alpha}\Bar{\beta}}({\bf k}^\prime,{\bf k})\frac{e^{i\left(({\bf k}^\prime-{\bf k})\cdot {\pmb \rho}-\xi\omega t)\right)}}{-\xi\hbar\omega-E_{\Bar{\beta} }(k)+E_{\Bar{\alpha}}( k^\prime) -i\delta}=0,\label{eq104}
\end{align}
where we use $\mathcal{G}^{\xi\dagger}_{\Bar{\alpha}\Bar{\beta}}({\bf k}^\prime,{\bf k})=\mathcal{G}^{-\xi}_{\Bar{\beta}\Bar{\alpha}}({\bf k},{\bf k}^\prime) $. In the second step, we change the variables ${\bf k}\leftrightarrow{\bf k}^\prime$, ${\bar \alpha}\leftrightarrow{\bar \beta}$, and $\xi\rightarrow-\xi$ for the second term.
With this relation, the spin-current density \eqref{spin_current_linear_response_simplfy} is simplified as 
\begin{align}
     {\bf J}_{s}^{(1)}({\pmb \rho})=\frac{\hbar^2}{4mA}\sum_{{\bf k}{\bf k}^\prime}\sum_{\xi=\pm}\sum_{\alpha,\beta=\{1,2\}}\left({\bf Q}_{\beta\alpha}({\bf k},{\bf  k}^\prime)\otimes({\bf k}+{\bf  k}^\prime)  \right)\mathcal{G}^\xi_{\alpha\beta}({\bf k}^\prime,{\bf k})\frac{e^{i\left(({\bf k}^\prime-{\bf k})\cdot {\pmb \rho}-\xi\omega t)\right)}}{\xi\hbar\omega-E_{\alpha}( k^\prime)+E_{\beta }(k) +i\delta}f(E_{\beta }(k))+{\rm H.c.}.
\label{spin_current_linear_response_simplfy_1}
 \end{align}

We then perform the summation over the wave vectors $\{{\bf k},{\bf k}'\}$ in Eq.~\eqref{spin_current_linear_response_simplfy_1} by the contour integral. In polar coordinates, ${\pmb \rho}=\rho\cos\varphi\hat{\bf y}+\rho\sin\varphi\hat{\bf z}$,  ${\bf k}=k\cos\theta_k\hat{\bf y}+k\sin\theta_k\hat{\bf z}$, and $e^{i{\bf k}\cdot{\pmb \rho}}=e^{ik\rho\cos(\varphi-\theta_k)}$. For the integral over $k^\prime$, when $(\theta_{k^\prime}-\varphi)\in(-\pi/2,\pi/2)$, $\cos(\theta_{k^\prime}-\varphi)>0$, so we close the integral contour in the upper complex plane, where the singularities are located at $(1/\hbar)\sqrt{{2m}(\mu+\sqrt{(E_{\beta}(k)+\xi\hbar\omega+i\delta)^2-\Delta_p^2})}\approx k_\omega^+(1+i\delta)$ and $-(1/\hbar)\sqrt{{2m}(\mu-\sqrt{(E_{\beta }(k)+\xi\hbar\omega+i\delta)^2-\Delta_p^2})}\approx -k_\omega^-(1-i\delta)$. Here, $k_\omega^{\pm }=({1}/{\hbar})\sqrt{{2m}(\mu\pm\sqrt{(E_{\beta }(k)+\xi\hbar\omega)^2-\Delta_p^2})}$. 
When $(\theta_{k^\prime}-\varphi)\in(\pi/2,3\pi/2)$, $\cos(\theta_{k^\prime}-\varphi)<0$, so we close the integral contour in the lower complex plane, where the singularities are located at $(1/\hbar)\sqrt{2m(\mu-\sqrt{(E_{\beta }(k)+\xi\hbar\omega+i\delta)^2-\Delta_p^2})}\approx k_\omega^-(1-i\delta)$ and $-(1/\hbar)\sqrt{2m(\mu+\sqrt{(E_{\beta }(k)+\xi\hbar\omega+i\delta)^2-\Delta_p^2})}\approx -k^+_\omega(1+i\delta)$.
When $\varphi\in[-\pi/2,\pi/2)$, we divide the integral region $\theta_{k^\prime}\in[0,\pi]$ into two parts $[0, \pi/2+\varphi]$ and $(\pi/2+\varphi,\pi]$ and calculate Eq.~\eqref{spin_current_linear_response_simplfy_1} as  
\begin{align}
{\bf J}_{s}^{(1)}({\pmb \rho})
   &=\frac{1}{8\pi}\sum_{{\bf k}}\sum_{\xi=\pm}\sum_{\alpha\beta}\frac{- i(E_{\beta }(k)+\xi\hbar\omega)}{\sqrt{(E_{\beta }(k)+\xi\hbar\omega)^2-\Delta^2_p}}f(E_{\beta }(k))e^{i[-k\rho\cos(\theta_{k}-\varphi)-\xi\omega t]}\nonumber\\
&\times\bigg{[}\int^{\varphi+\pi/2}_{\varphi-\pi/2}d\theta_{k^\prime}({\bf k}+{\bf k}^+_\omega)\otimes{\bf Q}_{\beta\alpha}\left({\bf k},{\bf k}^+_\omega\right)\mathcal{G}^{\xi}_{\alpha\beta}({\bf k}^+_\omega,{\bf k})e^{i[k^+_\omega \rho\cos(\theta_{k^\prime}-\varphi)-k\rho\cos(\theta_{k}-\varphi)-\xi\omega t]}\nonumber\\
  &+\int^{\varphi+\pi/2}_{\varphi-\pi/2}d\theta_{k^\prime}({\bf k}-{\bf k}^-_\omega)\otimes{\bf Q}_{\beta\alpha}\left({\bf k},-{\bf k}^-_\omega\right)\mathcal{G}^{\xi}_{\alpha\beta}(-{\bf k}^-_\omega,{\bf k})e^{i[-k^-_\omega \rho\cos(\theta_{k^\prime}-\varphi)-k\rho\cos(\theta_{k}-\varphi)-\xi\omega t]}\bigg{]},\label{eq106}
\end{align}
in which ${\bf k}^{+}_\omega=k_\omega^{+}(\cos\theta_{k^\prime},\sin\theta_{k^\prime})$ and ${\bf k}^{-}_\omega=k_\omega^{-}(\cos\theta_{k^\prime},\sin\theta_{k^\prime})$.

For the one-dimensional case, e.g., the exchange magnetic field is uniform along the $\hat{\bf y}$-direction with the Fourier components 
${\bf H}^\xi_{\rm ex}({\bf k})\rightarrow 2\pi\delta(k_z){\bf H}^\xi_{\rm ex}(k_y)$, the coupling matrix 
\begin{align}
\mathcal{G}^{\xi}_{\alpha\beta}({\bf k}^\prime,{\bf k})\rightarrow\frac{2\pi}{A}\delta\left(k^\prime\sin\theta_{k^\prime}-k_z\right)\left[{\bf H}^{\xi}_{\rm ex}(k^\prime\cos\theta_{k^\prime}-k_y)\cdot\pmb{\cal S}_{\alpha\beta}({\bf k}^\prime,{\bf k})\right].
\label{one-demonsional coupling matrix }
\end{align}
In this case, when away from the source, 
\begin{align}
&{\bf J}^{(1)}(y)=\frac{i}{4A}\sum_{ k_y,k_z}\sum_{\xi=\pm}\sum_{\alpha,\beta=1,2}\frac{-(E_{\beta}(k)+\xi\hbar\omega)}{\sqrt{(E_{\beta}(k)+\xi\hbar\omega)^2-\Delta^2_p}}f(E_{\beta}(k))\nonumber\\
  &\times\bigg{[}\frac{1}{k^+_\xi}\left((k_y+k^+_\xi)\hat{\bf y}+2k_z\hat{\bf z}\right)\otimes{\bf Q}_{\beta\alpha}[(k_y,k_z),(k^+_\xi,k_z)]{\bf H}^\xi_{\rm ex}(k^+_\xi-k_y)\cdot\pmb{\cal S}_{\alpha\beta}[(k^+_\xi,k_z),(k_y,k_z)]e^{i[(k^+_\xi-k_y)y-\xi\omega t]}\nonumber\\
  &+\frac{1}{k^-_\xi}\left((k_y-k^-_\xi)\hat{\bf y}+2k_z\hat{\bf z}\right)\otimes{\bf Q}_{\beta\alpha}[(k_y,k_z),(-k^-_\xi,k_z)]{\bf H}_{\rm ex}^\xi(-k^-_\xi-k_y)\cdot\pmb{\cal S}_{\alpha\beta}[(-k^-_\xi,k_z),(k_y,k_z)] e^{i[-(k^-_\xi+k_y) y-\xi\omega t]}\bigg{]}\nonumber\\
  &+{\rm H.c.},
  \label{spin_current_linear_field}
\end{align}
where $k_\xi^{\pm}=\sqrt{(2m/\hbar^2)\left(\mu\pm\sqrt{(E_{\beta}(k)+\xi\hbar\omega)^2-\Delta_p^2}\right)-k^2_z}$.

\subsection{DC spin-current density}

In the nonlinear-response regime, the spin-current density \eqref{spin_current_operator} reads
\begin{align}
    {\bf J}_{s}^{(2)}({\pmb \rho})&=\frac{i\hbar^2}{8mA}\sum_{{\bf k}{\bf k}^\prime}\bigg{[}\left\langle\hat{\textbf{\emph a}}^\dagger_{\bf k}U^\dagger_{\bf k}({\pmb \rho},t)(\overleftarrow{\pmb\nabla}-\overrightarrow{\pmb{\nabla}})\tau_3\pmb{\cal S}U_{{\bf k}^\prime}({\pmb \rho},t) \hat{\textbf{\emph b}}^{(2)}_{\bf k^\prime}\right\rangle\nonumber\\
    &+ \left\langle\hat{\textbf{\emph b}}^{(2)\dagger}_{\bf  k}U^\dagger_{\bf  k}({\pmb \rho},t)(\overleftarrow{\pmb\nabla}-\overrightarrow{\pmb{\nabla}})\tau_3\pmb{\cal S}U_{{\bf k}^\prime}({\pmb \rho},t) \hat{\textbf{\emph a}}_{\bf  k^\prime}\right\rangle+\left\langle\hat{\textbf{\emph b}}^{(1)\dagger}_{\bf k}U^\dagger_{\bf  k}({\pmb \rho},t)(\overleftarrow{\pmb\nabla}-\overrightarrow{\pmb{\nabla}})\tau_3\pmb{\cal S}U_{{\bf k}^\prime}({\pmb \rho},t) \hat{\textbf{\emph b}}^{(1)}_{\bf k^\prime}\right\rangle\bigg{]}\nonumber\\
     &=\frac{\hbar^2}{8mA}\sum_{{\bf k}{\bf k}^\prime{\bf q}}\sum^4_{\alpha\beta\gamma=1}\sum_{\xi_1\xi_2}\Bigg{[}{\bf Q}_{\beta\alpha}({\bf k},{{\bf  k}^\prime})\otimes({\bf k}+{\bf k}^\prime)\frac{\mathcal{G}^{\xi_1}_{\alpha\gamma}({\bf k}^\prime,{\bf q})\mathcal{G}^{\xi_2}_{\gamma\beta}({\bf q},{\bf k})e^{i[({\bf k}^\prime-{\bf k})\cdot{\pmb \rho}-(\xi_1+\xi_2)\omega t]}f(E_{\beta}(k))}{\left(E_{\gamma}(q)-E_{\beta}(k)-\xi_2\hbar\omega-i\delta\right)\left(E_{\alpha}(k^\prime)-E_{\beta}(k)+(\xi_1+\xi_2)\hbar\omega-i\delta\right)}\nonumber\\
    &+{\bf Q}_{\beta\alpha}({\bf k},{{\bf  k}^\prime})\otimes{\bf k}\frac{\mathcal{G}^{\xi_2}_{\alpha\gamma}({\bf k}^\prime,{\bf q})\mathcal{G}^{-\xi_1}_{\gamma\beta}({\bf q},{\bf k})e^{i[({\bf k}^\prime-{\bf k})\cdot{\pmb \rho}+(\xi_1-\xi_2)\omega t]}f(E_{\gamma}(q))}{\left(\xi_2\hbar\omega-E_{\beta}(k)+E_{\gamma}(q)-i\delta\right)\left(\xi_1\hbar\omega- E_{\alpha}(k^\prime)+E_{\gamma}(q)+i\delta\right)}\Bigg{]}+{\rm H.c.}.
    \label{spin_current_density_second_order}
\end{align}
We focus on the pumped DC spin-current density, i.e., taking $\xi_1+\xi_2=0$ in the first term and  $\xi_1=\xi_2$ in the second term of Eq.~\eqref{spin_current_density_second_order}. We also disregard the optical transition between the particle and hole bands when $\hbar \omega< 2\Delta_p$.
By the particle-hole symmetry, we thereby express the DC spin-current density \textit{within the particle space} as
\begin{align}
{\bf J}^{\rm DC}_{s}({\pmb \rho})    
&=\frac{\hbar^2}{8mA}\sum_{{\bf k}{\bf k}^\prime{\bf q}}\sum_{\xi=\pm}\sum_{\alpha\beta\gamma=\{1,2\}}{\bf Q}_{\beta\alpha}({\bf k},{{\bf  k}^\prime})\otimes{\bf k}^\prime\mathcal{G}^{-\xi}_{\alpha\gamma}({\bf k}^\prime,{\bf q})\mathcal{G}^{\xi}_{\gamma\beta}({\bf q},{\bf k})e^{i({\bf k}^\prime-{\bf k})\cdot{\pmb \rho}}\nonumber\\
&\times\Bigg{[}\frac{1}{\left(E_{\beta}(k)-E_{\alpha}(k^\prime)+i\delta\right)}\left(\frac{2f(E_{\alpha}(k^\prime))-1}{E_{\gamma}(q)-E_{\alpha}(k^\prime)-\xi\hbar\omega+i\delta}+\frac{2f(E_{\beta}(k))-1}{E_{\beta}(k)-E_{\gamma}(q)+\xi\hbar\omega+i\delta}\right)\nonumber\\
&+\frac{2f(E_{\gamma}(q))-1}{\left(E_{\gamma}(q)-E_{\alpha}(k^\prime)-\xi\hbar\omega+i\delta\right)\left(E_{\gamma}(q)-E_{\beta}(k)-\xi\hbar\omega-i\delta\right)}\Bigg{]}+{\rm H.c.}.
\label{eq114}
\end{align}
With the on-shell approximation, the DC spin-current density \eqref{eq114} becomes 
\begin{align}
{\bf J}^{\rm DC}_{s}({\pmb{\rho}})    
&=\frac{\hbar^2}{4mA}\sum_{{\bf k}{\bf k}^\prime{\bf q}}\sum_{\xi=\pm}\sum_{\alpha\beta\gamma=\{1,2\}}\big({\bf Q}_{\beta\alpha}({\bf k},{{\bf  k}^\prime})\otimes{\bf k}^\prime\big)\mathcal{G}^{-\xi}_{\alpha\gamma}({\bf k}^\prime,{\bf q})\mathcal{G}^{\xi}_{\gamma\beta}({\bf q},{\bf k})e^{i({\bf k}^\prime-{\bf k})\cdot{\pmb \rho}}\nonumber\\ 
&\times\frac{f(E_{\beta}(k))-f(E_{\gamma}(q))}{\left(E_{\gamma}(q)-E_{\alpha}(k^\prime)-\xi\hbar\omega+i\delta\right)\left(E_{\beta}(k)-E_{\gamma}(q)+\xi\hbar\omega+i\delta\right)}+{\rm H.c.}.
\label{nonlinear_approximation_current}
\end{align}

We then calculate the DC spin current \eqref{nonlinear_approximation_current} by using the contour integral.
We first perform the integral over $k^\prime$.   When $(\theta_{k^\prime}-\varphi)\in(-\pi/2,\pi/2)$, $\cos(\theta_{k^\prime}-\varphi)>0$, so we close the integral contour in the upper complex plane for $k^\prime$, where the singularities are located at $(1/\hbar)\sqrt{2m(\mu+\sqrt{(E_{\gamma}(q)-\xi\hbar\omega+i\delta)^2-\Delta_p^2})}\approx q_\omega^+(1+i\delta)$ and $-(1/\hbar)\sqrt{2m(\mu-\sqrt{(E_{\gamma}(q)-\xi\hbar\omega+i\delta)^2-\Delta_p^2})}\approx -q_\omega^-(1-i\delta)$, where 
$q_\omega^{\pm}=(1/\hbar)\sqrt{2m(\mu\pm \sqrt{(E_{\gamma}(q)-\xi\hbar\omega)^2-\Delta_p^2})}$.
When $(\theta_{k^\prime}-\varphi)\in(\pi/2,3\pi/2)$, $\cos(\theta_{k^\prime}-\varphi)<0$, so we close the integral contour in the lower complex plane, where the singularities are located at $(1/\hbar)\sqrt{2m(\mu-\sqrt{(E_{\gamma}(q)-\xi\hbar\omega+i\delta)^2-\Delta_p^2})}\approx q_\omega^-(1-i\delta)$ and $-(1/\hbar)\sqrt{2m(\mu+\sqrt{(E_{\gamma}(q)-\xi\hbar\omega+i\delta)^2-\Delta_p^2})}\approx -q^+_\omega(1+i\delta)$. When $\varphi\in[-\pi/2,\pi/2)$, we divide the integral region $\theta_{k^\prime}\in[0,\pi]$ into $[0, \pi/2+\varphi]$ and $(\pi/2+\varphi,\pi]$ and calculate Eq.~\eqref{nonlinear_approximation_current} as 
\begin{align}
    {\bf J}^{\rm DC}_{s}({\pmb \rho})
  & =\frac{1}{8\pi}\sum_{{\bf k}{\bf q}}\sum_{\xi=\pm}\sum_{\alpha\beta\gamma}\frac{-i(E_{\gamma}(q)-\xi\hbar\omega)}{\sqrt{(E_{\gamma}(q)-\xi\hbar\omega)^2-\Delta^2_p}}\frac{f(E_{\beta}(k))-f(E_{\gamma}(q))}{E_{\beta}(k)-E_{\gamma}(q)+\xi\hbar\omega+i\delta}e^{-ik\rho\cos(\theta_{k}-\varphi)}\nonumber\\
&\times\bigg{[}\int^{\pi/2+\varphi}_0d\theta_{k^\prime}{{\bf q}^+_\omega}\otimes{\bf Q}_{\beta\alpha}\left({\bf k},{\bf q}^+_\omega\right)\mathcal{G}^{-\xi}_{\alpha\gamma}({\bf q}^+_\omega,{\bf q})\mathcal{G}^{\xi}_{\gamma\beta}({\bf q},{\bf k})e^{iq^+_\omega \rho\cos(\theta_{k^\prime}-\varphi)}\nonumber\\
  &-\int^{\pi/2+\varphi}_0d\theta_{k^\prime}{{\bf q}^-_\omega}\otimes{\bf Q}_{\beta\alpha}\left({\bf k},-{\bf q}^-_\omega\right)\mathcal{G}^{-\xi}_{\alpha\gamma}(-{\bf q}^-_\omega,{\bf q})\mathcal{G}^{\xi}_{\gamma\beta}({\bf q},{\bf k})e^{-iq^-_\omega \rho\cos(\theta_{k^\prime}-\varphi)}\nonumber\\
  &+\int_{\pi/2+\varphi}^\pi d\theta_{k^\prime}{{\bf q}^-_\omega}\otimes{\bf Q}_{\beta\alpha}\left({\bf k},{\bf q}^-_\omega\right)\mathcal{G}^{-\xi}_{\alpha\gamma}({\bf q}^-_\omega,{\bf q})\mathcal{G}^{\xi}_{\gamma\beta}({\bf q},{\bf k})e^{iq^-_\omega \rho\cos(\theta_{k^\prime}-\varphi)}\nonumber\\
  &-\int_{\pi/2+\varphi}^\pi d\theta_{k^\prime}{{\bf q}^+_\omega}\otimes{\bf Q}_{\beta\alpha}\left({\bf k},-{\bf q}^+_\omega\right)\mathcal{G}^{-\xi}_{\alpha\gamma}(-{\bf q}^+_\omega,{\bf q})\mathcal{G}^{\xi}_{\gamma\beta}({\bf q},{\bf k})e^{-iq^+_\omega \rho\cos(\theta_{k^\prime}-\varphi)}\bigg{]},
  \label{eq120}
\end{align}
in which ${\bf q}^{+}_\omega=q_\omega^{+}(\cos\theta_{k^\prime}\hat{\bf y}+\sin\theta_{k^\prime}\hat{\bf z})$ and ${\bf q}^{-}_\omega=q_\omega^{-}(\cos\theta_{k^\prime}\hat{\bf y}+\sin\theta_{k^\prime}\hat{\bf z})$. 
We then perform the integral over $k$. 
When $(\theta_{k}-\varphi)\in(-\pi/2,\pi/2)$, $\cos(\theta_{k}-\varphi)>0$, so we close the integral contour in the lower complex plane, where the singularities are located at $(1/\hbar)\sqrt{2m(\mu+\sqrt{(E_{\gamma}(q)-\xi\hbar\omega-i\delta)^2-\Delta_p^2})}\approx q^+_\omega(1-i\delta)$ and $-(1/\hbar)\sqrt{2m(\mu-\sqrt{(E_{\gamma}(q)-\xi\hbar\omega-i\delta)^2-\Delta_p^2})}\approx-q_\omega^-(1+i\delta)$.  When $(\theta_{k}-\varphi)\in(\pi/2,3\pi/2)$, $\cos(\theta_{k}-\varphi)<0$, so we close the integral contour in the upper complex plane, where the singularities are located at $(1/\hbar)\sqrt{2m(\mu-\sqrt{(E_{\gamma}(q)-\xi\hbar\omega-i\delta)^2-\Delta_p^2})}\approx q^-_\omega(1+i\delta)$ and $-(1/\hbar)\sqrt{2m(\mu+\sqrt{(E_{\gamma}(q)-\xi\hbar\omega-i\delta)^2-\Delta_p^2})}\approx-q^+_\omega(1-i\delta)$.
The same calculation applies when $\varphi\in[\pi/2, 3\pi/2)$. The DC spin current reads
\begin{align}
    {\bf J}^{\rm DC}_{s}({\pmb \rho})&=\frac{\hbar^2}{4m}\frac{A}{(2\pi)^2}\sum_{{\bf q}}\sum_{\xi=\pm}\sum_{\alpha\beta\gamma}\left(\frac{m(E_{\gamma}(q)-\xi\hbar\omega)}{\hbar^2\sqrt{(E_{\gamma}(q)-\xi\hbar\omega)^2-\Delta^2_p}}\right)^2\big[f(E_{\gamma}(q))-f(E_{\gamma}(q)-\xi\hbar\omega)\big]\nonumber\\
  &\times\bigg{[}\int^{\pi/2+\varphi}_{\varphi-\pi/2}d\theta_k d\theta_{k^\prime}{{\bf q}^+_\omega}\otimes{\bf Q}_{\beta\alpha}\left({\bf p}^+_\omega,{\bf q}^+_\omega\right)\mathcal{G}^{-\xi}_{\alpha\gamma}({\bf q}^+_\omega,{\bf q})\mathcal{G}^{\xi}_{\gamma\beta}({\bf q},{\bf p}^+_\omega)e^{i[q^+_\omega \rho\cos(\theta_{k^\prime}-\varphi)-q^+_\omega \rho\cos(\theta_k-\varphi)]}\nonumber\\
  &+\int^{\pi/2+\varphi}_{\varphi-\pi/2}d\theta_k d\theta_{k^\prime}({-{\bf q}^-_\omega})\otimes{\bf Q}_{\beta\alpha}\left({\bf p}^+_\omega,-{\bf q}^-_\omega\right)\mathcal{G}^{-\xi}_{\alpha\gamma}(-{\bf q}^-_\omega,{\bf q})\mathcal{G}^{\xi}_{\gamma\beta}({\bf q},{\bf p}^+_\omega)e^{i[-q^-_\omega \rho\cos(\theta_{k^\prime}-\varphi)-q^+_\omega \rho\cos(\theta_k-\varphi)]}\nonumber\\
  &+\int^{\pi/2+\varphi}_{\varphi-\pi/2}d\theta_k d\theta_{k^\prime}{{\bf q}^+_\omega}\otimes{\bf Q}_{\beta\alpha}\left(-{\bf p}^-_\omega,{\bf q}^+_\omega\right)\mathcal{G}^{-\xi}_{\alpha\gamma}({\bf q}^+_\omega,{\bf q})\mathcal{G}^{\xi}_{\gamma\beta}({\bf q},-{\bf p}^-_\omega)e^{i[q^+_\omega \rho\cos(\theta_{k^\prime}-\varphi)+q^-_\omega \rho\cos(\theta_k-\varphi)]}\nonumber\\
  &+\int^{\pi/2+\varphi}_{\varphi-\pi/2}d\theta_k d\theta_{k^\prime}({-{\bf q}^-_\omega})\otimes{\bf Q}_{\beta\alpha}\left(-{\bf p}^-_\omega,-{\bf q}^-_\omega\right)\mathcal{G}^{-\xi}_{\alpha\gamma}(-{\bf q}^-_\omega,{\bf q})\mathcal{G}^{\xi}_{\gamma\beta}({\bf q},-{\bf p}^-_\omega)e^{i[-q^-_\omega \rho\cos(\theta_{k^\prime}-\varphi)+q^-_\omega \rho\cos(\theta_k-\varphi)]}\bigg{]}\nonumber\\
  &+{\rm H.c.}.
  \label{spin_current_nonlinear_integration}
  \end{align}

For the one-dimensional exchange magnetic field,  the DC spin-current density is obtained by substituting  the coupling matrix  Eq.~\eqref{one-demonsional coupling matrix } into  Eq.~\eqref{spin_current_nonlinear_integration}.

\section{Spin-torque density}

\label{spin_torque_appendix}

\subsection{Spin-torque density at the equilibrium}

Substitution of  Eq.~\eqref{field_operator_scattering} into (\ref{spin_torque_density_operator}) leads to the spin-torque density at the equilibrium
\begin{align}
    \mathbf{T}^{(0)}_s({\pmb{\rho}})=\frac{1}{4iA}\sum_{{\bf k}{\bf k}^\prime}\left\langle\hat{\pmb a}^\dagger_{\bf k}U^\dagger_{\bf k}({\pmb{\rho}},t){{\bf Q}^p}({\bf k},{\bf k}^\prime)U_{{\bf k}^\prime}({\pmb{\rho}},t)\hat{\pmb a}_{{\bf  k}^\prime}\right\rangle=\frac{1}{4i}\sum_{{\bf k}\beta}{{\bf Q}^{p}_{\beta\beta}}({\bf k},{\bf k})f(E_{\beta}(k)),
\end{align}
in which the matrix
\begin{align}
    {{\bf Q}^p}({\bf k},{\bf k}^\prime)=\Delta_p\bigg{[}\left(\begin{array}{cccc}
         0&0&-{e^{i\theta_{\bf k}}}&0  \\
         0&0&0&e^{-i\theta_{\bf k}}\\
         -e^{-i\theta_{\bf k}}&0&0&0\\
         0&e^{i\theta_{\bf k}}&0&0
    \end{array}\right)\left(\begin{array}{cc}
      {\pmb\sigma}   &  \\
         & -{\pmb\sigma^\ast}
    \end{array}\right)+\left(\begin{array}{cc}
      {\pmb\sigma}   &  \\
         & -{\pmb\sigma}^\ast
    \end{array}\right)\left(\begin{array}{cccc}
         0&0&{e^{i\theta_{{\bf k}^\prime}}}&0\\
         0&0&0&-e^{-i\theta_{{\bf k}^\prime}}\\
         e^{-i\theta_{{\bf k}^\prime}}&0&0&0\\
         0&-e^{i\theta_{{\bf k}^\prime}}&0&0
    \end{array}\right)\Bigg{]}.
\end{align}
The equilibrium spin-torque density $\mathbf{T}^{(0)}_s({\pmb{\rho}})$ vanishes  since ${{\bf Q}^p}({\bf k},{\bf k}')=0$ when ${\bf k}={\bf k}'$.

\subsection{Spin-torque density in the linear-response regime }

The spin-torque density in the linear-response regime reads
\begin{align}
    {\bf T}^{(1)}_s({\pmb{\rho}})&=\frac{1}{4iA}\sum_{{\bf k}{\bf k}^\prime}\left\langle \hat{\pmb a}^\dagger_{{\bf k}}U^\dagger_{\bf k}({\pmb{\rho}},t){\bf Q}^p({\bf k},{\bf k}^\prime)U_{{\bf k}^\prime}({\pmb{\rho}},t) \hat{\pmb b}^{(1)}_{{\bf k}^\prime}\right\rangle+{\rm H.c.}\nonumber\\
    &= \frac{1}{4iA}\sum_{{\bf k}{\bf k}^\prime}\sum_{\xi=\pm}\sum^4_{\beta\alpha=1}{\bf Q}^p_{\beta\alpha}({\bf k},{\bf k}^\prime)\mathcal{G}^\xi_{\alpha\beta}({\bf k}^\prime,{\bf k})\frac{e^{i({\bf k}^\prime-{\bf k})\cdot{\pmb{\rho}}-i\xi\omega t}f(E_{\beta}(k))}{\xi\hbar\omega-E_{\alpha}(k^\prime)+E_{\beta}(k)+i\delta}+{\rm H.c.}   \nonumber\\
    &\approx\frac{1}{4iA}\sum_{{\bf k}{\bf k}^\prime}\sum_{\xi=\pm}\Bigg{(}\sum_{\beta\alpha=\{1,2\}}{\bf Q}^p_{\beta\alpha}({\bf k},{\bf k}^\prime)\mathcal{G}^\xi_{\alpha\beta}({\bf k}^\prime,{\bf k})\frac{e^{i({\bf k}^\prime-{\bf k})\cdot{\pmb{\rho}}-i\xi\omega t}f(E_{\beta}(k))}{\xi\hbar\omega-E_{\alpha}(k^\prime)+E_{\beta}(k)+i\delta}\nonumber\\
    &+\sum_{\bar\beta\bar\alpha=\{3,4\}}{\bf Q}^p_{\Bar{\beta}\Bar{\alpha}}({\bf k},{\bf k}^\prime)\mathcal{G}^\xi_{\Bar{\alpha}\Bar{\beta}}({\bf k}^\prime,{\bf k})\frac{e^{i({\bf k}^\prime-{\bf k})\cdot{\pmb{\rho}}-i\xi\omega t}f(E_{\Bar{\beta} }(k))}{\xi\hbar\omega-E_{\Bar{\alpha}} (k^\prime)+E_{\Bar{\beta} }(k)+i\delta}\Bigg{)}+{\rm H.c.}.
    \label{spin_torque_linear}
\end{align}
When $T\rightarrow 0$~K, the spin-torque density in the linear-response regime vanishes:
\begin{align}
    {\mathbf T}^{(1)}_s({\pmb{\rho}})&\rightarrow \frac{1}{4iA}\sum_{{\bf k}{\bf k}^\prime}\sum_{\bar\beta\bar\alpha=\{3,4\}}\sum_{\xi=\pm}{\bf Q}^p_{\Bar{\beta}\Bar{\alpha}}({\bf k},{\bf k}^\prime)\mathcal{G}^\xi_{\Bar{\alpha}\Bar{\beta}}({\bf k}^\prime,{\bf k})\frac{e^{i({\bf k}^\prime-{\bf k})\cdot{\pmb{\rho}}-i\xi\omega t}}{\xi\hbar\omega-E_{\Bar{\alpha}}( k^\prime)+E_{\Bar{\beta} }(k)+i\delta}+{\rm H.c.}\nonumber\\
    &=\frac{1}{4iA}\sum_{{\bf k}{\bf k}^\prime}\sum_{\bar\beta\bar\alpha=\{3,4\}}\sum_{\xi=\pm}{\bf Q}^p_{\Bar{\beta}\Bar{\alpha}}({\bf k},{\bf k}^\prime)\mathcal{G}^\xi_{\Bar{\alpha}\Bar{\beta}}({\bf k}^\prime,{\bf k})\frac{e^{i({\bf k}^\prime-{\bf k})\cdot{\pmb{\rho}}-i\xi\omega t}}{\xi\hbar\omega-E_{\Bar{\alpha}}( k^\prime)+E_{\Bar{\beta} }(k)+i\delta}\nonumber\\
    &+\frac{1}{4iA}\sum_{{\bf k}{\bf k}^\prime}\sum_{\bar\beta\bar\alpha=\{3,4\}}\sum_{\xi=\pm}{\bf Q}^p_{\Bar{\alpha}\Bar{\beta}}({\bf k}^\prime,{\bf k})\mathcal{G}^{-\xi}_{\Bar{\beta}\Bar{\alpha}}({\bf k},{\bf k}^\prime)\frac{e^{i({\bf k}-{\bf k}^\prime)\cdot{\pmb{\rho}}+i\xi\omega t}}{\xi\hbar\omega-E_{\Bar{\alpha}}( k^\prime)+E_{\Bar{\beta} }(k)-i\delta}
    =0,
\end{align}
in which ${\bf Q}^{p\dagger}_{\bar\beta\bar\alpha}({\bf k},{\bf k}^\prime)=-{\bf Q}^{p}_{\bar\alpha\bar\beta}({\bf k}^\prime,{\bf k})$ is used. 
With this property, Eq.~\eqref{spin_torque_linear} is reduced to
\begin{align}
    {\mathbf T}^{(1)}_s({\pmb{\rho}})=&\frac{1}{2iA}\sum_{{\bf k}{\bf k}^\prime}\sum_{\xi=\pm}\sum_{\beta\alpha=\{1,2\}}{\bf Q}^p_{\beta\alpha}({\bf k},{\bf k}^\prime)\mathcal{G}^\xi_{\alpha\beta}({\bf k}^\prime,{\bf k})\frac{e^{i({\bf k}^\prime-{\bf k})\cdot{\pmb{\rho}}-i\xi\omega t}}{\xi\hbar\omega-E_{\alpha}(k^\prime)+E_{\beta}(k)+i\delta}f(E_{\beta}(k))+{\rm H.c.}.
    \label{spin_torque_linear_simplfy}
\end{align}
By the contour integral, Eq.~\eqref{spin_torque_linear_simplfy} becomes 
\begin{align}
      {\mathbf T}^{(1)}_s({\pmb{\rho}})&=\frac{1}{4\pi}\sum_{{\bf k}}\sum_{\xi=\pm}\sum_{\alpha\beta}\frac{- m(E_{\beta}(k)+\xi\hbar\omega)}{\hbar^2\sqrt{(E_{\beta}(k)+\xi\hbar\omega)^2-\Delta^2_p}}f(E_{\beta}(k))\nonumber\\
&\times\bigg{[}\int^{\varphi+\pi/2}_{\varphi-\pi/2}d\theta_{k^\prime}{\bf Q}^p_{\beta\alpha}\left({\bf k},{\bf k}^+_\omega\right)\mathcal{G}^{\xi}_{\alpha\beta}({\bf k}^+_\omega,{\bf k})e^{i[k^+_\omega \rho\cos(\theta_{k^\prime}-\varphi)-k\rho\cos(\theta_{k}-\varphi)-\xi\omega t]}\nonumber\\
  &+\int^{\varphi+\pi/2}_{\varphi-\pi/2}d\theta_{k^\prime}{\bf Q}^p_{\beta\alpha}\left({\bf k},-{\bf k}^-_\omega\right)\mathcal{G}^{\xi}_{\alpha\beta}(-{\bf k}^-_\omega,{\bf k})e^{i[-k^-_\omega \rho\cos(\theta_{k^\prime}-\varphi)-k\rho\cos(\theta_{k}-\varphi)-\xi\omega t]}\bigg{]}+{\rm H.c.},
\end{align}
in which ${\bf k}^{+}_\omega=k_\omega^{+}(\cos\theta_{k^\prime},\sin\theta_{k^\prime})$ and ${\bf k}^{-}_\omega=k_\omega^{-}(\cos\theta_{k^\prime},\sin\theta_{k^\prime})$.
Driven by the one-dimensional exchange field, with the coupling constant Eq.~\eqref{one-demonsional coupling matrix }, the spin-torque density when away from the source
\begin{align}
     &{\mathbf T}^{(1)}(y)=\frac{1}{2A}\sum_{ k_y,k_z}\sum_{\xi=\pm}\sum_{\alpha\beta=1}^2\frac{-m(E_{\beta}(k)+\xi\hbar\omega)}{\hbar^2\sqrt{(E_{\beta}(k)+\xi\hbar\omega)^2-\Delta^2_p}}f(E_{\beta}(k))\nonumber\\
     &\times\bigg{[}{\bf Q}^p_{\beta\alpha}[(k_y,k_z),(k^+_\xi,k_z)]{\bf H}_{\rm ex}^\xi(k^+_\xi-k_y)\cdot\pmb{\cal S}_{\alpha\beta}[(k^+_\xi,k_z),(k_y,k_z)]e^{i(k^+_\xi y-k_yy-\xi\omega t)}\nonumber\\
  &+{\bf Q}^p_{\beta\alpha}[(k_y,k_z),(-k^-_\xi,k_z)]{\bf H}_{\rm ex}^\xi(-k^-_\xi-k_y)\cdot\pmb{\cal S}_{\alpha\beta}[(-k^-_\xi,k_z),(k_y,k_z)]e^{i(-k^-_\xi y-k_yy-\xi\omega t)}\bigg{]}+{\rm H.c.}. 
\end{align}

\subsection{DC spin-torque density}

In the nonlinear-response regime, the spin-torque density 
\begin{align}
    \mathbf{T}^{(2)}_s({\pmb{\rho}})=&\frac{1}{4iA}\sum_{{\bf k}{\bf k}^\prime{\bf q}}\sum_{\xi_1\xi_2=\pm}\sum^4_{\alpha\beta\gamma=1 }\bigg{(}{\bf Q}^p_{\beta\alpha}({\bf k},{\bf k}^\prime)\frac{{\mathcal G}^{\xi_1}_{\alpha\gamma}({\bf k}^\prime-{\bf q}){\mathcal G}^{\xi_2}_{\gamma\beta}({\bf q}-{\bf k})e^{i({\bf k}^\prime-{\bf k})\cdot{\pmb{\rho}}-i(\xi_1+\xi_2)\omega t}f(E_{\beta}(k))}{(E_{\gamma}(q)-E_{\beta}(k)-\xi_2\hbar\omega-i\delta)(E_{\alpha}(k^\prime)-E_{\beta}(k)-(\xi_1+\xi_2)\hbar\omega -i\delta)}\nonumber\\
    +&\tilde{\bf Q}^{p}_{\beta\alpha}({\bf k},{\bf k}^\prime)\frac{{\mathcal G}^{\xi_2}_{\alpha\gamma}({\bf k}^\prime-{\bf q}){\mathcal G}^{-\xi_1}_{\gamma\beta}({\bf q}-{\bf k})e^{i({\bf k}^\prime-{\bf k})\cdot{\pmb{\rho}}+i(\xi_1-\xi_2)\omega t}f(E_{\gamma}(q))}{(E_{\gamma}(q)-E_{\alpha}(k^\prime)+\xi_2\hbar\omega+i\delta)(\xi_1\hbar\omega -E_{\beta}(k)+E_{\gamma}(q)-i\delta)}\bigg{)}+{\rm H.c.},
    \end{align}
    in which the matrix elements
\begin{align}
    \tilde{\bf Q}^{p}_{\beta\alpha}({\bf k},{\bf k}^\prime)=\Delta_p\phi^\dagger_\beta({\bf k})\left(\begin{array}{cccc}
         0&0&-{e^{i\theta_{\bf k}}}&0  \\
         0&0&0&e^{-i\theta_{\bf k}}\\
         -e^{-i\theta_{\bf k}}&0&0&0\\
         0&e^{i\theta_{\bf k}}&0&0
    \end{array}\right)\left(\begin{array}{cc}
      {\pmb\sigma}   &  \\
         & -{\pmb\sigma^\ast}
    \end{array}\right)\phi_\alpha({\bf k}^\prime).
\end{align}
 The DC component
    \begin{align}
    \mathbf{T}^{\rm DC}_s({\pmb{\rho}})
   &=\frac{1}{2iA}\sum_{{\bf k}{\bf k}^\prime{\bf q}}\sum_{\xi=\pm}\sum_{\alpha\beta\gamma=1}^2\overline{\bf Q}^{p}_{\beta\alpha}({\bf k},{{\bf  k}^\prime})\mathcal{G}^{-\xi}_{\alpha\gamma}({\bf k}^\prime,{\bf q})\mathcal{G}^{\xi}_{\gamma\beta}({\bf q},{\bf k})e^{i({\bf k}^\prime-{\bf k})\cdot{\pmb{\rho}}}\nonumber\\
&\times\frac{f(E_{\beta}(k))-f(E_{\gamma}(q))}{\left(E_{\gamma}(q)-E_{\alpha}(k^\prime)-\xi\hbar\omega+i\delta\right)\left(E_{\beta}(k)-E_{\gamma}(q)+\xi\hbar\omega+i\delta\right)}+{\rm H.c.}, 
\label{spin_torque_nonlinear} 
\end{align}
in which the matrix elements 
\begin{align}
   \overline{\bf Q}^{p}_{\beta\alpha}({\bf k},{{\bf  k}^\prime})=\Delta_p\phi^\dagger_\beta({\bf k})\left(\begin{array}{cc}
      {\pmb\sigma}   &  \\
         & -{\pmb\sigma^\ast}
    \end{array}\right)\left(\begin{array}{cccc}
         0&0&{e^{i\theta_{{\bf k}^\prime}}}&0  \\
         0&0&0&-e^{-i\theta_{{\bf k}^\prime}}\\
         e^{-i\theta_{{\bf k}^\prime}}&0&0&0\\
         0&-e^{i\theta_{{\bf k}^\prime}}&0&0
    \end{array}\right)\phi_\alpha({\bf k}^\prime).
\end{align}
In the absence of the quasiparticle population at zero temperature, the DC spin-torque density $\mathbf{T}_s^{\rm DC}({\pmb{\rho}})$ vanishes.

For the one-dimensional exchange field, with the coupling constant Eq.~\eqref{one-demonsional coupling matrix }, the DC spin-torque density
\begin{align}
    &\mathbf{T}^{\rm DC}_s(y)
   =\frac{1}{8\pi^2iA}\sum_{{\bf q}}\sum_{\xi=\pm}\sum_{\alpha\beta\gamma=1}^2\int dk^\prime_y dk_y\overline{\bf Q}^{p}_{\beta\alpha}[( k_y,q_z)(k^\prime_y,q_z)]e^{i(k^\prime_y-k_y)y}\nonumber\\
   &\times\left({\bf H}_{\rm ex}^{-\xi}(k_y^\prime-q_y)\cdot\pmb{\cal S}_{\alpha\gamma}[(k^\prime_y,q_z),(q_y,q_z)]\right)\left({\bf H}_{\rm ex}^{\xi}(q_y-k_y)\cdot\pmb{\cal S}_{\gamma\beta}[(q_y,q_z),(k_y,q_z)]\right)\nonumber\\
&\times\frac{f(E_{\beta} (k_y,q_z))-f(E_{\gamma}(q))}{\left(E_{\gamma}(q)-E_{\alpha }(k_y^\prime,q_z)-\xi\hbar\omega+i\delta\right)\left(E_{\beta} (k_y,q_z)-E_{\gamma}(q)+\xi\hbar\omega+i\delta\right)}+{\rm H.c.}.
\label{spin torque outside the field}
\end{align}
\end{widetext} 
By the contour integral over the associated expression and numerical calculation, we can obtain the DC spin-torque density induced by the exchange interaction.

\end{appendix}

\end{document}